\documentclass[prb,a4paper,showpacs,
twocolumn]{revtex4-1}
\usepackage{amsmath}
\usepackage{amssymb}
\usepackage{amsfonts}
\usepackage{graphicx}
\usepackage[usenames,dvipsnames]{color}
\usepackage{bm}
\usepackage{ulem}

\DeclareMathOperator{\Tr}{Tr}

\DeclareMathOperator{\sgn}{sgn} \DeclareMathOperator{\Imag}{Im}
\DeclareMathOperator{\Real}{Re} 
 
\DeclareMathOperator{\erfi}{erfi}
\DeclareMathOperator{\erf}{erf}

\DeclareMathOperator{\Prob}{Prob}

 1

\begin{document}

\maxdeadcycles=2000
\title{Spin fluctuations in quantum dots}

\author{A.U. Sharafutdinov, D.S. Lyubshin, and I.S. Burmistrov}
\affiliation{L.D. Landau Institute for Theoretical Physics RAS,
Kosygina street 2, 119334 Moscow, Russia}
\affiliation{Moscow
Institute of Physics and Technology, 141700 Moscow, Russia}


\begin{abstract}
We explore the static longitudinal and dynamic transverse spin susceptibilities in quantum dots and nanoparticles
within the framework of the Hamiltonian that extends the universal Hamiltonian to the case of uniaxial anisotropic exchange. For the limiting cases of Ising and Heisenberg exchange interactions we ascertain how fluctuations of single-particle levels affect the Stoner instability in quantum dots. We reduce the problem to the statistics of extrema of a certain Gaussian process.  We prove that, in spite possible strong randomness of the single-particle levels, the spin susceptibility and all its moments diverge simultaneously at the point which is determined by the standard criterion of the Stoner instability involving the mean level spacing only.
\end{abstract}


\maketitle


\section{Introduction}

The physics of quantum dots continuously attracted a lot of experimental and theoretical interest.~\cite{Alhassid2000,Wiel,ABG,Hanson,Ullmo2008} Within the assumption that the Thouless energy ($E_{\rm Th}$) is much larger than mean single-particle level spacing ($\delta$), $E_{\rm Th}/\delta \gg 1$, an effective zero-dimensional Hamiltonian has been derived. \cite{KAA} In this so-called  universal Hamiltonian the electron-electron interaction that involves a set of matrix elements in the single-particle basis is reduced to just three parameters: the charging energy ($E_c$), the ferromagnetic exchange ($J>0$) and the interaction in the Cooper channel. The single particle energies are random quantities with Wigner-Dyson statistics. Thus the universal Hamiltonian provides a convenient framework for the theoretical description of quantum dots.

The charging energy (typically $E_c\gg \delta$) restricts the probability of real electron tunneling through a quantum dot at low temperatures $T\ll E_c$. \cite{CB} This phenomenon of the Coulomb blockade leads to suppression of the tunneling density of states in quantum dots at low temperatures \cite{KamenevGefen1996,SeldmayrLY}. It was also understood that a small enough exchange interaction $J\lesssim \delta/2$ is important for a quantitative description of the experiments on low temperature ($T\lesssim \delta$) transport through quantum dots fabricated in a two-dimensional electron gas. \cite{QDLowT}
For a quantum dot of size $L \gg \lambda_F$ ($\lambda_F$ stands for the Fermi wave length) the exchange interaction can be estimated by bulk value of the Fermi-liquid interaction parameter ($F_0^\sigma$): $J/\delta = -
F_0^\sigma$. As it is well-known, strong enough exchange interaction in bulk materials leads to a Stoner instability at $F_0^\sigma=-1$ and a corresponding quantum phase transition between a paramagnet and a ferromagnet. In quantum dots it is possible to realize an interesting situation in which the ground state has a finite total spin.  \cite{KAA} In the case of  the equidistant single-particle spectrum it occurs for $\delta/2 \lesssim J<\delta$. As $J$ increases towards $\delta$, the total spin in the ground state increases and at $J=\delta$ all electrons in a quantum dot become spin polarized. This phenomenon of mesoscopic Stoner instability is specific to finite size systems and disappears in the thermodynamic limit $\delta \to 0$. Due to the entanglement of the charge and spin degrees of freedom in the universal Hamiltonian, the mesoscopic Stoner instability affects the electron transport through a quantum dot. For example, it leads to an additional nonmonotonicity of the energy dependence of the tunneling density of states \cite{KiselevGefen,BGK1,BGK2} and to the enhancement of the shot noise. \cite{Koenig2012} The Cooper channel interaction in the description within the universal Hamiltonian framework is responsible for superconducting correlations in quantum dots. \cite{SCinQD} We shall assume throughout the paper that the Cooper channel interaction is repulsive and, therefore, omit it. \cite{KAA} We also neglect corrections to the universal Hamiltonian due to the fluctuations in the matrix elements of the electron-electron interaction. \cite{Altshuler1997,Mirlin1997} They are small in the regime $\delta/E_{\rm Th}\ll 1$ but lead to interesting physics beyond the universal Hamiltonian. \cite{Ullmo2008}

In the presence of a spin-orbit coupling the description of a quantum dot in the framework of the universal Hamiltonian breaks down. Even for a weak spin-orbit coupling (large spin-orbit length, $\lambda_{SO} \gg L$)
fluctuations of the matrix elements of the electron-electron interaction cannot be neglected in spite of the condition $\delta/E_{\rm Th}\ll 1$. \cite{AlhassidSO,SOinQD} For a quantum dot in a two-dimensional electron gas the orbital degrees of freedom are coupled to in-plane components of the spin. Then in the regime $(\lambda_{SO}/L)^2 \gg (E_{\rm Th}/\delta)(L/\lambda_{SO})^4 \gg 1$ the low energy description is again possible in terms of the universal Hamiltonian but with the Ising exchange interaction ($J_z>0$). \cite{AlhassidSO,AF2001} In this case mesoscopic Stoner instability is absent for the equidistant single-particle spectrum. \cite{KAA} As a consequence, the tunneling density of states is almost independent of $J_z$ while the longitudinal spin susceptibility $\chi_{zz}$ is independent of $T$ as in a clean Fermi liquid. \cite{KiselevGefen,Boaz}

The experiments on tunneling spectra in nanometer-scale ferromagnetic nanoparticles revealed the presence of an exchange interaction with significant anisotropy. \cite{Gueron1999} The simplest model which allows to explain the main features of experimentally measured excitation spectra of ferromagnetic nanoparticles resembles the universal Hamiltonian with uniaxial anisotropy in exchange interaction. \cite{Canali} Such modification of exchange interaction can arise due to shape, surface, or bulk magnetocrystalline anisotropy. In addition, in the presence of spin-orbit scattering the anisotropic part of the exchange interaction can experience large mesoscopic fluctuations. \cite{UB2005, BG2005} The alternative reason for appearance of anisotropy in the exchange interaction in quantum dots is the presence of ferromagnetic leads. \cite{Misiorny2013}

The universal Hamiltonian with an anisotropic exchange interaction (albeit it is not microscopically justified) is interesting on its own as the simplest model interpolating between the cases of the Heisenberg and Ising exchange interactions. Since in the latter case there is no mesoscopic Stoner instability for the equidistant single-particle spectrum, it is interesting to understand how it disappears as the exchange develops anisotropy. Does the spin of the ground state vanish continuously or discontinuously as the anisotropy increases? For the Ising exchange interaction transverse dynamical spin susceptibility $\chi_\perp(\omega)$ is nontrivial. Its imaginary part is odd in frequency with maxima and minima at $\omega = \pm \omega_{\rm ext}$, respectively. \cite{Boaz} In the case of the Heisenberg exchange $\Imag\chi_\perp(\omega)$ reduces to a delta-function. But how does this reduction occur with decrease in anisotropy?

In low dimensions $d \leqslant 2$ interaction and disorder can induce a transition between paramagnetic and ferromagnetic phases at a finite temperature $T$. \cite{Finkelstein,KamenevAndreev1998}
In $d=3$ the Stoner instability can be promoted by disorder and occurs at smaller values of exchange interaction \cite{Nayak2005}. In the universal Hamiltonian the disorder remains in randomness of the single-particle levels. As it is known, \cite{KAA,BGK2} level fluctuations affect the temperature dependence of the average static spin susceptibility $\chi_{zz}$ in the case of the Heisenberg exchange. In the case of the Ising exchange the role of disorder is even more dramatic. For the equidistant single-particle levels $\chi_{zz}$ is temperature independent. Due to level fluctuations the average spin susceptibility acquires a Curie type $T$-dependence dominating at low enough $T$ and for $\delta-J_z\ll\delta$. \cite{KAA} In this regime of strong (with respect to the small distance $\delta-J_z \ll \delta$ to the average position of the Stoner instability at $J_z=\delta$) level fluctuations   a quantum dot is in the paramagnetic phase on average but it can be fully spin-polarized for a particular realization of the single-particle levels. These fully spin-polarized realizations should affect the tails of the distribution functions for $\chi_{zz}$ and dynamical transverse spin susceptibility $\chi_\perp(\omega)$, but how exactly? Can it be possible that at zero temperature the level fluctuations shift the position of the Stoner instability from its average position, $J_z=\delta$, and lead to the existence of a finite temperature transition between the paramagnetic and the ferromagnetic phases in quantum dots? Of course, the very same questions can be asked for the case of the Heisenberg exchange.

In this paper we address these questions within the universal Hamiltonian framework extended to the case of exchange interaction with uniaxial anisotropy. We compute the temperature and magnetic field dependence of
the static longitudinal spin susceptibility $\chi_{zz}$ for equidistant single-particle spectrum. Except the case of the Ising exchange it always has a non-zero temperature-dependent contribution of
Curie type ($1/T$) or of $1/\sqrt{T}$ type. This indicates that destruction of the mesoscopic Stoner instability by uniaxial anisotropy is not abrupt. For equidistant single-particle levels we also compute the transverse spin susceptibility. It always has a maximum and a minimum whose positions tend to zero frequency with decrease of anisotropy.  We show that at low temperatures and for $\delta-J_z\ll\delta$ the statistical properties of the longitudinal spin susceptibility (both for the Ising and Heisenberg exchanges) are determined by the statistics of the extrema of a certain Gaussian process with a drift. This random process resembles locally a fractional Brownian motion with the Hurst exponent $H=1-\epsilon$ where $\epsilon \to 0$. We recall that the fractional Brownian motion with the Hurst exponent $H$ is the Gaussian process $B_H(t)$ with zero mean $\overline{B_H(t)}=0$ and the two-point correlation function $\overline{[B_H(t)-B_H(t^\prime)]^2} = |t-t^\prime|^{2H}$. We rigorously prove that  in the case of Ising (Heisenberg) exchange all moments of static longitudinal spin susceptibility $\chi_{zz}$ are finite for $J_z<\delta$ ($J<\delta$). For the Ising exchange we argue also that all moments of dynamic transverse spin susceptibility $\chi_\perp(\omega)$ do not diverge for  $J_z<\delta$. We estimate the tail of the complementary cumulative distribution function for $\chi_{zz}$ for both Ising and Heisenberg exchange interactions. We demonstrate that the average static longitudinal spin susceptibility $\chi_{zz}$ has nonmonotonous dependence on magnetic field in the case of Ising exchange. Our results mean that the level fluctuations do not shift the Stoner instability from its average position and do not induce a finite temperature transition between the paramagnetic and the ferromagnetic phases.

The outline of the paper is as follows. In Sec. \ref{sec:formalism} we introduce the model Hamiltonian, derive exact analytical expressions for the corresponding grand canonical partition function and longitudinal static spin susceptibility. In Sec. \ref{sec:EqSpectrum} we analyze the temperature and magnetic field dependence of longitudinal static spin susceptibility in the case of equidistant single-particle spectrum and anisotropic exchange interaction. In Sec. \ref{sec:LF:IC} we present a detailed analysis of the effect of level fluctuations on the longitudinal static spin susceptibility for the cases of Ising and Heisenberg exchange interactions. In Sec.
\ref{sec:trss} we compute the transverse dynamical spin susceptibility in the case of equidistant single-particle spectrum and anisotropic exchange interaction and analyze the effect of level fluctuations in the case of Ising exchange interaction. We conclude the paper with summary of the main results  and discussion of how our predictions can be experimentally verified (Sec. \ref{sec:dc}). Some of the results were published in a brief form in Ref. [\onlinecite{LSB2014}].

\section{Hamiltonian and partition function \label{sec:formalism}}

\subsection{Hamiltonian}
\label{AUH}

We consider the following Hamiltonian with direct Coulomb and anisotropic exchange interactions:
\begin{equation}
H =H_0+H_{C}+H_{S}.
\label{ham}
\end{equation}
The noninteracting Hamiltonian,
\begin{equation}
H_0=\sum_{\alpha,\sigma}\epsilon_{\alpha,\sigma} a^{\dag}_{\alpha\sigma}a_{\alpha\sigma}   ,
\end{equation}
is given as usual in terms of the single-particle creation ($a^{\dag}_{\alpha\sigma}$) and annihilation ($a_{\alpha\sigma}$) operators. It involves the spin-dependent ($\sigma=\pm$) single-particle energy levels $\epsilon_{\alpha,\sigma}$. In what follows, we assume that
they depend on applied magnetic field $B$ via the Zeeman splitting, $\epsilon_{\alpha,\sigma} = \epsilon_\alpha+g_L\mu_B B \sigma/2$.  Here $g_L$ and $\mu_B$ stand for the Land\'e g-factor and the Bohr magneton, respectively. The charging interaction part of the Hamiltonian,
\begin{equation}
H_C=E_c(\hat{n}-N_0)^2 ,
\label{Hc}
\end{equation}
describes the direct Coulomb interaction in a quantum dot in the zero-dimensional approximation, $E_{\rm Th}/\delta\gg 1$. Here
\begin{equation}
\hat{n} =\sum_{\alpha} n_\alpha = \sum_{\alpha,\sigma}a^{\dag}_{\alpha,\sigma}a_{\alpha,\sigma}
\end{equation}
denotes the particle number operator, and $N_0$ is the background charge. The term
\begin{equation}
H_S=-J_{\perp}(\hat{S}_x^2+\hat{S}_y^2)-J_z\hat{S}_z^2 ,
\label{Hs}
\end{equation}
represents the anisotropic exchange interaction within the QD. The total spin operator
\begin{equation}
\hat{\bm{S}}=\frac{1}{2} \sum_{\sigma\sigma^\prime} a^\dag_{\alpha\sigma}\bm{\sigma}_{\sigma\sigma^\prime}a_{\alpha\sigma^\prime}
\end{equation}
is defined in terms of the standard Pauli matrices $\bm{\sigma}$.

In the case of isotropic Heisenberg exchange, $J_\perp=J_z$, the Hamiltonian \eqref{ham} reduces to the universal Hamiltonian which describes a quantum dot in the limit $E_{\rm Th}/\delta \gg 1$. \cite{KAA} In this case the single-particle levels $\epsilon_\alpha$ are random. Their statistics (in the absence of magnetic field, $B=0$) is described by the orthogonal Wigner-Dyson ensemble. The Hamiltonian \eqref{ham} with the Ising exchange, $J_\perp=0$, and $B=0$ can be used for description of lateral quantum dots with spin-orbit coupling. \cite{AF2001,AlhassidSO} In this case the statistics of $\epsilon_\alpha$ is described by the unitary Wigner-Dyson ensemble.

\subsection{Exact expression for the grand canonical partition function}

The grand canonical partition function for the Hamiltonian \eqref{ham} is defined as $Z= \Tr e^{-\beta H+\beta \mu \hat n }$ ($\mu$ denotes the chemical potential). It can be found by using the following trick. Let us separate $H_S$  into the Heisenberg and Ising parts:
\begin{equation}
H_S=-J_\perp \hat{\bm{S}}^2-(J_z-J_\perp) \hat{S}_z^2 .
\end{equation}
Then the time evolution operator in the imaginary time can be rewritten as 
\begin{align}
\label{teo}
e^{-\tau {H}_S} = & \frac{\sqrt{\tau}}{2\sqrt{\pi |J_z-J_\perp|}}\int\limits_{-\infty}^{\infty} d\mathcal{B}\, \exp\left (-\frac{\tau\mathcal{B}^2}{4|J_z-J_\perp|}\right )  \notag \\
& \hspace{2cm} \times e^{\tau J_\perp \hat{\bm{S}}^2-\eta \mathcal{B}\hat{S_z}}  ,
\end{align}
where $\eta=\sqrt{\sgn(J_z-J_\perp)}$. The exponent in the second line of Eq. \eqref{teo} indicates that the grand canonical partition function for the Hamiltonian \eqref{ham} can be found in two steps. At first, one can use well-known results for the partition function for the case of isotropic exchange and effective magnetic field $B + \eta \mathcal{B}/(g_L\mu_B)$. \cite{AlhassidRupp,BGK2} Secondly, one needs to integrate over the effective magnetic field $\mathcal{B}$ with the kernel given in the first line of Eq. \eqref{teo}. Thus we obtain  the following exact result for the grand canonical partition function of Hamiltonian \eqref{ham}:
\begin{gather}
Z(b) = \sum\limits_{n_\uparrow,n_\downarrow} Z_{n_\uparrow}Z_{n_\downarrow}
e^{-\beta E_c(n-N_0)^2+\beta J_\perp m(m+1)+\beta\mu n} \notag \\
\times \sgn \left (2m+1\right )  \sum\limits_{l=-|m+1/2|+1/2}^{|m+1/2|-1/2}  e^{\beta(J_z-J_\perp)l^2 - \beta b l}.
\label{eqZb1}
\end{gather}
Here $b=g_L \mu_B B/2$. The integers $n_\uparrow$ and $n_\downarrow$ represent the number of spin-up and spin-down
electrons, respectively. The total number of electrons is $n = n_\uparrow+n_\downarrow$, and $m = (n_\uparrow - n_\downarrow)/2$. We note that for a configuration with given $n_\uparrow$ and $n_\downarrow$ electrons the total spin equals $S = |m+1/2|-1/2$. The integers $l$ denote $z$ projection of the total spin $\bm{S}$. The factors $Z_{n_\uparrow}$ and $Z_{n_\downarrow}$ are canonical partition functions for $n_\uparrow$ and $n_\downarrow$ noninteracting spinless electrons, respectively. The canonical partition function takes into account the contributions from the single-particle energies and is given by Darwin-Fowler integral:
\begin{equation}
Z_n=\int\limits_0^{2\pi} \frac{d\theta}{2\pi}\, e^{-i n\theta}\prod_{\gamma}\left (1+e^{i\theta-\beta\epsilon_\gamma}\right ).
\end{equation}

For the Heisenberg exchange interaction, $J_\perp=J_z$ our result \eqref{eqZb1} coincides with the result known in the literature. \cite{AlhassidRupp,BGK1,BGK2} In the case of purely Ising exchange interaction, $J_\perp=0$, our result \eqref{eqZb1} agrees with the result obtained in Ref. [\onlinecite{Boaz}]. We note that the result \eqref{eqZb1} can be also derived directly from Hamiltonian \eqref{ham} with the help of Wei-Norman-Kolokolov transformation (see Appendix \ref{WNK-der}).

In order to analyze the exact result \eqref{eqZb1} for the grand canonical partition function, it will be convenient to use the following completely equivalent integral representation:
\begin{gather}
Z(b)=\frac{e^{-\beta J_\perp/4}}{2 \pi \sqrt{J_\perp |J_z-J_\perp|}}\int_{-\infty}^{\infty}dh d\mathcal{B}\,
e^{-\frac{h^2}{\beta J_\perp}} e^{-\frac{\beta(b+\eta \mathcal{B})^2}{4 J_\perp}}
\notag \\
\times \frac{\sinh (h) \sinh \bigl ((b+\eta \mathcal{B})h/J_\perp\bigr )}{\sinh \bigl (\beta(b+\eta \mathcal{B})/2\bigr )}\sum_{k\in \mathbb{Z}} e^{-\beta E_c(k-N_0)^2} \notag \\ \times  e^{-\frac{\beta \mathcal{B}^2}{4|J_z-J_\perp|}}
\int\limits_{-\pi}^{\pi}\frac{d\phi_0}{2\pi}  e^{i\phi_0 k}
\prod_{\sigma}e^{-\beta \Omega_0(\mu-i\phi_0 T +\sigma h T)} .
\label{IR}
\end{gather}
The grand canonical partition function for non-interacting spinless electrons is defined in a standard way
\begin{equation}
\Omega_0(\mu)=-T \ln \prod_{\gamma}\left (1+e^{-\beta (\epsilon_{\gamma}-\mu)}\right ).
\end{equation}
The variables $\phi_0$ and $h$ have the meaning of the zero-frequency Matsubara components of an electric potential and a magnetic field which can be used to decouple the direct Coulomb \cite{KamenevGefen1996} and exchange interaction \cite{Boaz,Saha2012} terms, respectively.

\subsection{The longitudinal spin susceptibility \label{sec:longsusc}}

The general expressions \eqref{eqZb1} and \eqref{IR} for the grand partition function $Z$ allow us to extract the results for the longitudinal spin susceptibility:
\begin{equation}
\chi_{zz}(T,b)=T\frac{\partial^2}{\partial b^2}\ln Z .
\end{equation}
It is worthwhile to mention that in zero magnetic field one can use the equivalent formula
\begin{equation}
\chi_{zz}(T,b=0)=\partial \ln Z/\partial J_z
\label{eq:chizzb0}
\end{equation}
to simplify calculations.  As it is well-known,\cite{KamenevGefen1996,EfetovTscherisch} at $T\gg \delta$ (the regime we are interested in) we can perform integration over $\phi_0$ in Eq. \eqref{IR} in the saddle-point approximation. Then the grand canonical partition function is factorized into two multipliers:
\begin{equation}
Z = Z_C Z_S, \label{ZCZS}
\end{equation}
where
\begin{equation}
Z_C = \sqrt{\frac{\beta\Delta}{4\pi}} \sum_{n\in\mathbb{Z}} e^{-\beta E_c(n-N_0)^2 +\beta (\mu-\mu_n) n -2 \beta \Omega_0(\mu_n)} , \label{ZC1}
\end{equation}
describes the effect of charging energy. Here $\mu_n$ is the solution of the saddle-point equation $n = -2 \partial \Omega_0(\mu)/\partial \mu$ and
\begin{equation}
\Delta^{-1} = - \frac{\partial^2\Omega_0(\mu)}{\partial \mu^2} \Bigl |_{\mu=\mu_n}
\end{equation}
stands for the thermodynamic density of states at the Fermi level. We note that in the regime $T\ll E_c$ (which we are interested in)  one can approximate $\mu_n$ by $\tilde{\mu} = \mu_{N_0}$. The term
\begin{align}
Z_S=& \frac{e^{-\beta J_\perp/4}}{2\pi \sqrt{J_\perp |J_z-J_\perp|}}\int_{-\infty}^{\infty}dh d\mathcal{B}\, e^{-\frac{1}{4 \beta J_\perp}[ 4h^2+\beta^2(b+\eta \mathcal{B})^2]} \notag \\
&\hspace{1cm}\times \frac{\sinh (h) \sinh \bigl ((b+\eta \mathcal{B})h/J_\perp\bigr )}{\sinh \bigl (\beta(b+\eta \mathcal{B})/2\bigr )} e^{-\frac{\beta\mathcal{B}^2}{4 |J_z-J_\perp|}]} \notag \\
&\hspace{1cm} \times  \prod_{\sigma}e^{\beta \Omega_0(\tilde\mu)-\beta \Omega_0(\tilde\mu+h \sigma/\beta)}\label{ZS1}
\end{align}
describes the contribution due to exchange interaction. The function
\begin{gather}
 \beta\sum_\sigma [\Omega_0(\tilde{\mu})-\Omega_0(\tilde{\mu}+ h\sigma/\beta)]
 = \int\limits_{-\infty}^\infty dE\, \nu_0(E)\,  \notag \\
 \times \ln \left [ 1+\frac{\sinh^2(h/2)}{\cosh^2(E/2T)}\right ]
 \end{gather}
that appears in Eq.~\eqref{ZS1} depends on a particular realization of the single-particle spectrum via the single-particle density of states $\nu_0(E)=\sum_\alpha \delta(E+\tilde{\mu}-\epsilon_\alpha)$. Provided
$h^2\ll \exp(\beta\tilde{\mu})$, we can write
\begin{equation}
\beta\sum_\sigma \Bigl [\Omega_0(\tilde{\mu})-\Omega_0(\tilde{\mu}+ h\sigma/\beta)\Bigr ] = \frac{h^2}{\beta \delta}-V(h) ,\label{ZZeq2}
\end{equation}
where
 \begin{equation}
V(h) = -\int\limits_{-\infty}^\infty dE\, \delta\nu_0(E)\,  \ln \left [ 1+\frac{\sinh^2(h/2)}{\cosh^2(E/2T)}\right ] .\label{Vh_Def}
\end{equation}
Here $\delta \nu_0(E)$ stands for the deviation of the single-particle density of states $\nu_0(E)$ from its average (over realizations of the single-particle spectrum) value:
$1/\delta = \overline{1/\Delta}=\overline{\nu_0(E)}$.

The charging energy contribution $Z_C$ is independent of the magnetic field and therefore does not affect the spin susceptibility. We note that the normalization is such that $Z_S=1$ for $b=J_\perp=J_z=0$. In what follows we will discuss $Z_S$ only.

\section{The longitudinal spin susceptibility: Equidistant single-particle spectrum \label{sec:EqSpectrum}}

We start our analysis from the case of the equidistant single-particle spectrum, i.e. we completely neglect the effect of level fluctuations (we set $V(h)$ in Eq.\eqref{ZZeq2} to zero). We discuss the role of level fluctuations in Sec. \ref{sec:LF:IC}.

\subsection{The case of an easy axis: $J_z\geqslant J_\perp$}

Using the integral representation \eqref{ZS1} and \eqref{ZZeq2} we can perform integration over $h$ and find
\begin{align}
Z_S &=  \left ( \frac{\delta}{\delta-J_z}\right )^{1/2} e^{\frac{\beta J_\perp^2}{4(\delta-J_\perp)}} e^{-\frac{\beta b^2}{4(J_z-J_\perp)}} \notag \\
& \times  \frac{1}{2}  \sum_{p=\pm} F_1\left ( \frac{\delta}{\delta-J_\perp}+\frac{p b}{J_z-J_\perp},\sqrt{\beta J_*} \right) .
\label{eqZZb1}
\end{align}
Here $J_*={(\delta-J_\perp)(J_z-J_\perp)}/{(\delta-J_z)}$ is the energy scale specific for the anisotropic problem that interpolates between $0$ (for $J_z=J_\perp$) and $\delta J_z/(\delta-J_z)$ (for $J_\perp=0$).
The function $F_1(x,y)$ is defined as follows
\begin{equation}
F_1(x,y) = \int\limits_{-\infty}^{\infty}\frac{dt}{\sqrt{\pi}}\frac{\sinh(x y t)}{\sinh (y t)}\,e^{-t^2} .
\label{eq:F1}
\end{equation}

Using Eq. \eqref{eq:chizzb0}, the zero field longitudinal spin susceptibility can be written as
\begin{align}
\chi_{zz}(T) & = \frac{1}{2(\delta-J_z)} + \frac{1}{2} \left ( \frac{\delta-J_\perp}{\delta-J_z}\right)^2 \notag  \\
& \times  \frac{\partial}{\partial J_*} \ln F_1\left (\frac{\delta}{\delta-J_\perp},\sqrt{\beta J_*} \right).
\label{eq:chizz0gen}
\end{align}

At high temperatures $T \gg \max\left\{\delta,\frac{\delta^2(J_z-J_\perp)}{(\delta-J_\perp)(\delta-J_z)}\right \}$,
the result  \eqref{eq:chizz0gen} for the zero-field longitudinal static spin susceptibility can be simplified (cf. Eq. \eqref{eq:App:F1:ZZF1r1}). Then we obtain
\begin{equation}
\chi_{zz}(T) = \frac{1}{2(\delta-J_z)} + \frac{\beta}{12}\frac{(2\delta-J_\perp)J_\perp}{(\delta-J_z)^2} .
\label{eq:chizzT0}
\end{equation}

Away from the isotropic case ($J_z=J_\perp$) a set of temperature intervals with different temperature behavior of the longitudinal spin susceptibility exists. Below we use the asymptotic result \eqref{eq:App:F1:ZZF1r2} from Appendix \ref{App:F1:ZZF1}.  At temperatures $\max\left \{\delta,\frac{\delta(J_z-J_\perp)}{(\delta-J_z)}\right \} \ll T \ll \frac{\delta^2(J_z-J_\perp)}{(\delta-J_\perp)(\delta-J_z)}$, we find
\begin{equation}
\chi_{zz}(T) = \frac{1}{2(\delta-J_z)} + \frac{\beta}{4}\frac{\delta^2}{(\delta-J_z)^2} .
\label{eq:chizzT1}
\end{equation}
For the temperature range $\max\left \{\delta,\frac{(\delta-J_\perp)(J_z-J_\perp)}{(\delta-J_z)}\right \} \ll T \ll \frac{\delta(J_z-J_\perp)}{(\delta-J_z)}$, we obtain
\begin{equation}
\chi_{zz}(T) = \frac{1}{2(\delta-J_z)} +\frac{\beta}{4}\frac{J_\perp^2}{(\delta-J_z)^2} .
\label{eq:chizzT2}
\end{equation}
If the temperature is within the interval  $\max\left \{\delta, \frac{J_\perp^2(J_z-J_\perp)}{(\delta-J_\perp)(\delta-J_z)}\right \}\ll  T \ll \frac{(\delta-J_\perp)(J_z-J_\perp)}{(\delta-J_z)}$, the zero field longitudinal static spin susceptibility becomes
\begin{equation}
\chi_{zz}(T) = \frac{1}{2(\delta-J_z)} + \frac{1}{2\sqrt\pi} \frac{J_\perp\sqrt{\beta J_*}}{(\delta-J_z)(J_z-J_\perp)} .
\label{eq:chizzT3}
\end{equation}
Finally, for the lowest temperature range $\delta \ll T \ll \min\left \{\frac{J_\perp^2(J_z-J_\perp)}{(\delta-J_\perp)(\delta-J_z)},\frac{(\delta-J_\perp)(J_z-J_\perp)}{(\delta-J_z)}\right \}$  we find (cf. Eq. \eqref{eq:App:F1:ZZF1r3})
\begin{equation}
\chi_{zz}(T) = \frac{1}{2(\delta-J_z)} +\frac{\beta}{4}\frac{J_\perp^2}{(\delta-J_z)^2} .
\label{eq:chizzT4}
\end{equation}

We mention that $\chi_{zz}$ consists of two contributions (see Eqs. \eqref{eq:chizzT0}-\eqref{eq:chizzT2} and \eqref{eq:chizzT4}): the one which resembles the Fermi-liquid result for spin susceptibility, $\propto 1/(\delta-J_z)$, and the other which is of Curie type,  $\propto \beta \delta^2/(\delta-J_z)^2$. Such behavior is illustrated in Fig. \ref{fig:plot_chi_zz} where the dependence of longitudinal spin susceptibility \eqref{eq:chizz0gen} on temperature and $J_z$ at a fixed ratio $J_\perp/\delta$ is shown. We emphasize that longitudinal spin susceptibility diverges at $J_z=\delta$ regardless of the value of $J_\perp$.

To understand the origin of such interesting behavior of the zero field longitudinal spin susceptibility it is useful to rewrite Eq. \eqref{ZS1} in terms a series form again:
\begin{align}
Z_S&=
\sqrt{\frac{\beta \delta}{\pi}} e^{\frac{\beta J_\perp^2}{4(\delta-J_\perp)}
}\sum_{S_z=-\infty}^{\infty}\sum_{S=|S_z|}^{\infty} \Biggl (
e^{-\beta(\delta-J_\perp)(S-\frac{J_\perp}{2(\delta-J_\perp)})^2}\notag \\ &- e^{-\beta(\delta-J_\perp)(S+1+\frac{J_\perp}{2(\delta-J_\perp)})^2}\Biggr )e^{\beta(J_z-J_\perp)S_z^2} .
\label{Z_reason}
\end{align}
Here we used the following result
\begin{equation}
Z_{n_\uparrow}Z_{n_\downarrow} \approx \sqrt{\frac{\beta \delta}{4\pi}}
e^{-\beta\mu_n n - 2 \beta \Omega_0(\mu_n)}
 \sqrt{\frac{\beta \delta}{\pi}}e^{-\beta \delta m^2}
\label{EQm2}
\end{equation}
which is valid provided the following conditions hold: $\delta\ll T$ and $n\gg |m|$ (see Appendix \ref{App:ZnZn}).

\begin{figure}[t]
\centerline{ \includegraphics[width=7cm]{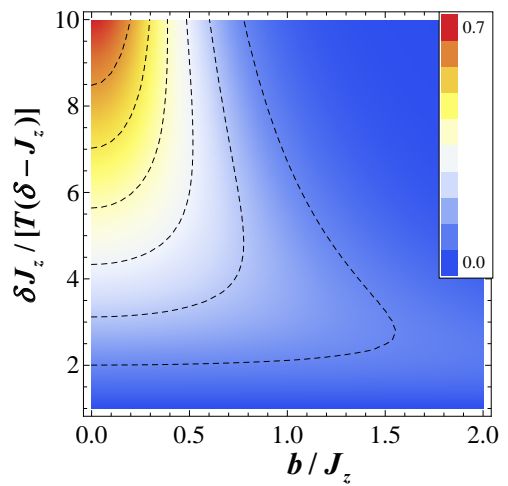}}
\caption{(Color online) Dependence of the relative correction $[2(\delta-J_z)\chi_{zz}-1]$ to the Fermi-liquid-like result on dimensionless magnetic field and inverse temperature, $b/J_z$ and $\delta J_z/(\delta-J_z)T$. We chose $J_z=0.94 \delta$ and $J_\perp=0.3 \delta$.}
\label{fig:chi_zz_on_b_2D}
\end{figure}

In the case of large temperatures $T\gg J_\perp^2/(\delta-J_z)$ our results \eqref{eq:chizzT0}-\eqref{eq:chizzT4} imply the Fermi-liquid behavior of $\chi_{zz}(T)$. In this temperature range all terms except the first one with $S=|S_z|$ in the sum over $S$ in Eq. \eqref{Z_reason} cancel each other. Then we find
\begin{eqnarray}
Z_S= \sqrt{\frac{\beta \delta}{\pi}} \sum_{S_z=-\infty}^{\infty}e^{-\beta(\delta-J_z)S_z^2}  = \left (\frac{\delta}{\delta-J_z}\right )^{1/2}
\end{eqnarray}
and, consequently, $\chi_{zz}(T)=1/[2(\delta-J_z)]$. This result implies that the average value of $S_z^2$ is of the order of $1/[2\beta(\delta-J_z)] \gg 1$ regardless of $J_\perp$. At the same time the average value of the squared total spin $\bm{S}^2$ is of the order of $1/[2\beta(\delta-J_z)]+1/[\beta(\delta-J_\perp)]$. Therefore, at $J_\perp\lesssim J_z$ the total spin strongly fluctuates in all three directions so that $\bm{S}^2 \approx 3 S_z^2$ whereas for $J_\perp \ll J_z$ the total spin fluctuates along the $z$ axis only so that $\bm{S}^2 \approx S_z^2$.

We mention the unusual (inverse square-root) temperature dependence of the longitudinal spin susceptibility in Eq. \eqref{eq:chizzT3}. However, the result \eqref{eq:chizzT3} is valid in a temperature range that exists only if $J_\perp\ll J_z\lesssim \delta$. Then the restrictions for the temperature become $\max\{\delta, J_\perp^2/(\delta-J_z\} \ll T \ll \delta^2/(\delta-J_z)$. Therefore we can use the arguments from the previous paragraph. In order to explain the $\sqrt{\beta}$ dependence of $\chi_{zz}$, one needs to perform the perturbation expansion in $J_\perp |S_z| \sim J_\perp /\sqrt{\beta(\delta-J_z)}$ for Eq. \eqref{Z_reason}.

At low temperatures $T\ll J_\perp^2/(\delta-J_z)$ our results \eqref{eq:chizzT0}-\eqref{eq:chizzT2} and \eqref{eq:chizzT4} imply a Curie type longitudinal spin susceptibility. In this case the second term in brackets in the right hand side of Eq. \eqref{Z_reason} can be neglected. The sum over $S$ can be estimated by the integral which is dominated by $S\sim |S_z|$. Then we find
\begin{align}
Z_S=\sqrt{\frac{\beta \delta}{\pi}} \, e^{\frac{\beta J_\perp^2}{4(\delta-J_z)}} \sum_{S_z=-\infty}^{\infty} \frac{e^{-\beta(\delta-J_z)(|S_z|-\frac{J_\perp}{2(\delta-J_z)})^2}}{2\beta(\delta-J_\perp)|S_z|-\beta J_\perp}  .
\label{eq:ZsLowApprox}
\end{align}
This estimate yields the typical value of $|S_z| = J_\perp/[2(\delta-J_z)]$ and, thus, the Curie type behavior of the longitudinal spin susceptibility: $\chi_{zz}=\beta |S_z|^2 = \beta J_\perp^2/[2(\delta-J_z)]^2$. Therefore, at relatively low temperatures $\delta \ll T\ll J_\perp^2/(\delta-J_z)$ the configuration with a non-zero total spin $S=|S_z| = J_\perp/[2(\delta-J_z)]$ gives the main contribution to the thermodynamic quantities.

\begin{figure}[t]
\includegraphics[width=7cm]{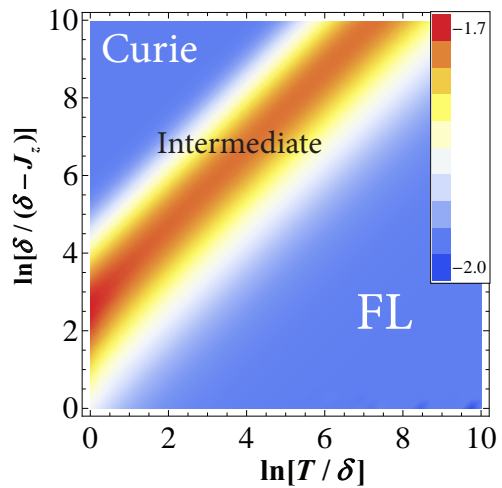}
\caption{(Color online) Dependence of $-\frac{d \ln [\chi_{zz}-\frac{\delta}{2(\delta-J_z)}]}{d (\delta-J_z)}$ on $\ln \frac{T}{\delta}$ and $\ln \frac{\delta}{\delta-J_z}$ for $J_\perp=0.3\delta$. {In the left upper region the Curie-like behavior dominates the Fermi-liquid-like result. In the right lower region the Curie-like correction to the Fermi-liquid-like result, $\Bigl[\chi_{zz}-\frac{\delta}{2(\delta-J_z)} \Bigr]\propto \frac{1}{(\delta-J_z)^2}$ is small. Red region corresponds to the intermediate regime in which there is a correction $\Bigl[\chi_{zz}-\frac{\delta}{2(\delta-J_z)}\Bigr]\propto \frac{1}{T^{1/2}(\delta-J_z)^{3/2}}$ to the Fermi-liquid result due to transverse degrees of freedom.}}
\end{figure}

For small magnetic fields $b\ll \delta(J_z-J_\perp)/(\delta-J_\perp)$ the longitudinal spin susceptibility $\chi_{zz}(T,b)$ can be well approximated by the zero field result. For larger magnetic fields $b\gg \delta(J_z-J_\perp)/(\delta-J_\perp)$ there are two regions of temperature with different behavior. In the range of temperatures $b(\delta-J_\perp)/(\delta-J_z) \ll T \ll b \delta/(\delta-J_z)$ the longitudinal spin susceptibility becomes linear in temperature (cf. Eq. \eqref{eq:App:F1:ZZF1r1}):
\begin{equation}
\chi_{zz}(T,b) = \frac{1}{2(\delta-J_z)} + \frac{T}{b^2} .
\label{LbEA0}
\end{equation}
At higher temperatures $T\gg b \delta/(\delta-J_z)$ the temperature dependence of the longitudinal spin susceptibility saturates:
\begin{equation}
\chi_{zz}(T,b) = \frac{1}{2(\delta-J_z)}.
\label{LbEA}
\end{equation}

In the limit of large magnetic fields the ground state energy for the configuration with the total spin projection $S_z$ is equal to $(\delta-J_z)S_z^2-bS_z$. Thus the projection of the total spin in the ground state is $S_z=b/[2(\delta-J_z)]$. It allows us to estimate the longitudinal spin susceptibility as $\chi_{zz}=d S_z /db=1/[2(\delta-J_z)]$ in agreement with Eq. \eqref{LbEA}.

\subsection{The case of the easy plane ($J_z < J_\perp$)}

Using the integral representation \eqref{ZS1} and \eqref{ZZeq2}, we integrate over $h$ and obtain
\begin{align}
Z_S &=  \left ( \frac{\delta}{\delta-J_z}\right )^{1/2} e^{\frac{\beta (J_\perp^2+b^2)}{4(\delta-J_\perp)}}  \int \limits_{-{\pi}/{2}}^{{\pi}/{2}}
\frac{dt}{\sqrt{\pi}}\, e^{-\frac{t^2}{\beta|J_*|}+\frac{i b t}{\delta-J_\perp}} \notag \\
& \times  \frac{\sinh\Bigl (\frac{\delta(\beta b+2it)}{2(\delta-J_\perp)} \Bigr )}{\sqrt{\beta|J_*|} \sinh\bigl ( \frac{\beta b+2it}{2}\bigr )}
 \vartheta_3\Bigl (e^{-\frac{\pi^2}{\beta(J_\perp-J_z)}},\frac{i \pi t}{\beta(J_\perp-J_z)}\Bigr ) .
 \label{eq:Zs:theta}
\end{align}
Here $\vartheta_3(q,z) = \sum_m q^{m^2} e^{2i m z}$ stands for the Jacobi theta function. Since $T\gg \delta \geqslant J_\perp-J_z$, the Jacobi theta function $\vartheta_3$ becomes equal to unity. Then for $b=0$ we find
\begin{align}
Z &=  \left ( \frac{\delta}{\delta-J_z}\right )^{1/2} e^{\frac{\beta J_\perp^2}{4(\delta-J_\perp)}}  F_2\left ( \frac{\delta}{\delta-J_\perp}, \sqrt{\beta|J_*|} \right ) ,
\end{align}
where
\begin{equation}
F_2(x,y) = \int \limits_{-\pi/2y}^{\pi/2y} \frac{dt}{\sqrt\pi}  \frac{\sin(xyt)}{\sin(yt)}\, e^{-t^2} .
\label{eq:F2}
\end{equation}

At temperatures $T \gg \max\left\{\delta,\frac{\delta^2(J_\perp-J_z)}{(\delta-J_\perp)(\delta-J_z)}\right \}$, with the help of Eq. \eqref{eqnF2asymp} we obtain that the longitudinal spin susceptibility is given by Eq. \eqref{eq:chizzT0}.
In the temperature range $\delta \ll T \ll \frac{\delta^2(J_\perp-J_z)}{(\delta-J_\perp)(\delta-J_z)}$, the behavior of $\chi_{zz}$ is described by Eq. \eqref{eq:chizzT1}. In the case of an easy plane anisotropy the interplay between Fermi-liquid and Curie-like temperature dependencies of the longitudinal spin susceptibility can be explained in exactly  the same way as it was done for the case of an easy axis anisotropy.

The longitudinal static spin susceptibility is almost insensitive to the presence of a small magnetic field
$b\ll \delta(J_\perp-J_z)/(\delta-J_z)$. In the opposite case $b\gg \delta(J_\perp-J_z)/(\delta-J_z)$, one
can neglect $t$ in the $\sinh$'s arguments in Eq. \eqref{eq:Zs:theta}. Then at $b\gg \delta(J_\perp-J_z)/(\delta-J_z)$ we find
\begin{equation}
Z_S = \left ( \frac{\delta}{\delta-J_z}\right )^{1/2} e^{\frac{\beta J_\perp^2}{4(\delta-J_\perp)}+\frac{\beta b^2}{4(\delta-J_z)}} \,
 \frac{\sinh\frac{\delta \beta b}{2(\delta-J_\perp)}}{\sinh \frac{\beta b}{2}} .
 \label{eq:Zs:EP:simplified}
\end{equation}
The result \eqref{eq:Zs:EP:simplified} implies that for magnetic fields in the range $(\delta-J_\perp)T/\delta\ll b \ll T$ the longitudinal spin susceptibility is described by Eq. \eqref{LbEA0} whereas for $b\gg T$, $\chi_{zz}$ is given by Eq. \eqref{LbEA}.

\section{The longitudinal spin susceptibility: The effect of level fluctuations \label{sec:LF:IC}}

As it was explained above, the Hamiltonian \eqref{ham} describes a quantum dot in the zero-dimensional limit for the Ising and Heisenberg exchange interactions only. Therefore, it is reasonable to study the effect of level fluctuations on the results obtained above for $J_\perp=0$ and $J_\perp=J_z$. We start with the case of Ising exchange.

\subsection{The Ising exchange}

To simplify the general result \eqref{ZS1} in the case of the Ising exchange, it is convenient to make a change of variable $\mathcal{B} \to \mathcal{B} -2 h (J_z-J_\perp)T/J_z$, to take the limit $J_\perp\to 0$,  and then to integrate over $\mathcal{B}$. Thus we find
\begin{equation}
Z_S = \left (\frac{\delta}{\delta-J_z}\right)^{1/2} e^{\frac{\beta b^2}{4(\delta-J_z)}}
\Xi\left (\frac{b}{J_z},\frac{\beta J_z\delta}{\delta-J_z}\right) ,
\label{eq:ZS:Xi}
\end{equation}
where
\begin{equation}
\Xi(x,y) = \int\limits_{-\infty}^\infty \frac{dh}{\sqrt{\pi}}
e^{-h^2-V(h\sqrt{y}+x y/2)} .
\label{eq:Xi:def}
\end{equation}
The information on fluctuations of single-particle levels is encoded in the even random function $V(h)$ via the density of states (see Eq. \eqref{Vh_Def}). We remind that the single-particle density of states $\nu_0$(E) has
non-Gaussian statistics.~\cite{Mehta} However, for $\max\{|h|,T/\delta\}\gg 1$  the function
$V(h)$ is a Gaussian random variable with zero mean value.~\cite{Mehta} The two-point correlation function of $V$ can be written as follows (see Appendix \ref{app:Lh}):
\begin{gather}
\overline{V(h_1)V(h_2)}
= \sum_{\sigma=\pm} L(h_1+\sigma h_2)- 2 L(h_1) - 2L(h_2) ,  \notag \\
L(h) = \frac{2}{\pi^2\bm{\beta}} \int\limits_0^{|h|} dt \, t\,  \left [\Real \psi\left (1+\frac{i t}{2\pi}\right )+\gamma\right ] .
\label{corrVV}
\end{gather}
Here  $\psi(z)$ is the Euler digamma function and $\gamma=-\psi(1)$ is the Euler--Mascheroni constant. In the case of the Ising exchange, the parameter $\bm{\beta}$ in Eq. \eqref{corrVV} is equal to $\bm{\beta}=2$ since the energy levels $\epsilon_\alpha$ in the Hamiltonian \eqref{ham} are described by the unitary Wigner-Dyson ensemble (class A).~\cite{AF2001} The asymptotics of $L(h)$ are as follows:~\cite{BGK2}
\begin{equation}
L(h) = \frac{4}{\bm{\beta}}\left (\frac{h}{2\pi}\right )^2\begin{cases}
\frac{\zeta(3)}{2} \left (\frac{h}{2\pi}\right )^2 - \frac{\zeta(5)}{3}\left (\frac{h}{2\pi}\right )^4,\, &\frac{|h|}{2\pi}\ll 1,\\
\ln \frac{|h|}{2\pi}+\gamma-\frac12, \, & \frac{|h|}{2\pi}\gg 1 .
\end{cases} \label{Sassymp}
\end{equation}

\subsubsection{Perturbation expansion for $\overline{\chi}_{zz}$}

\begin{figure}[t]
\centerline{\includegraphics[width=6cm]{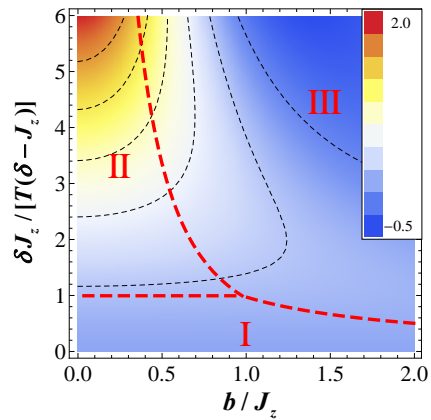}}
\caption{(Color online) Different regions of behavior of the relative correction to $\overline{\chi}_{zz}$ due to fluctuations for the case of Ising exchange in the plane of dimensionless magnetic field and inverse temperature, $b/J_z$ and $\delta J_z/T(\delta-J_z)$. Note that in our analysis we assume $T\gg \delta$.}
\label{FigureRegions}
\end{figure}

According to Eq. \eqref{eq:ZS:Xi}, the average longitudinal spin susceptibility $\overline{\chi}_{zz}$ is determined by the quantity $\overline{\ln \Xi(x,y)}$. Although $V(h)$ is a Gaussian random variable, exact evaluation of $\overline{\ln \Xi(x,y)}$ for arbitrary values of $x$ and $y$ is a complicated problem. We start from the perturbation theory in the correlation function $\overline{V(h)V(h^\prime)}$. Expanding
expression~\eqref{eq:ZS:Xi} for $\Xi(x,y)$ to the second order in $V$ and performing the averaging of $\ln \Xi(x,y)$ with the help of Eq.~\eqref{corrVV}, we find
\begin{align}
\overline{\ln \Xi(x,y)} & = \int \limits_{0}^\infty \frac{du}{\sqrt{\pi}} \, e^{-u^2} \Bigl [ e^{-x^2 y/4}\cosh(u x\sqrt{y})L(2u\sqrt{y}) \notag \\
- & \bigl ( e^{-x^2 y/2} \cosh(u x\sqrt{2 y}) +1 \bigr )
L(u\sqrt{2 y})
\Bigr ] .
\label{AverLnXi1}
\end{align}
There exist four regions of different behavior of $\overline{\ln \Xi(x,y)}$. They are shown in Fig.~\ref{FigureRegions}. It is convenient to introduce the renormalized exchange $\bar{J}_z= \delta J_z/(\delta-J_z)$.

In the region I, $\bar{J}_z \max\{1,(b/J_z)\} \ll T$,  the arguments of  $\Xi(x,y)$ satisfy the condition $y\ll \min\{1,1/x\}$. The latter allows one to use the asymptotics of $L(h)$ for $|h|\ll1$ (see Eq.~\eqref{Sassymp}). Then we  find

\begin{equation}
\overline{\ln \Xi(x,y)} = \frac{3\zeta(3) y^2}{8\pi^4 \bm{\beta}} \Bigl [ 1 + y x^2  - \frac{5 \zeta(5) y}{2\pi^2\zeta(3)}\Bigl (1+\frac{3}{2} y x^2 + \frac{ y^2x^4}{6}\Bigr )\Bigr ] .\label{eq:Xi:regI}
\end{equation}

Hence we obtain the following result for the average longitudinal spin susceptibility at temperatures $T\gg \bar{J}_z \max\{1,(b/J_z)\}$:

\begin{eqnarray}
\overline{\chi}_{zz} &=& \frac{1}{2(\delta-J_z)} + \frac{3\zeta(3)}{4\pi^4 \bm{\beta}}  \frac{\delta^3 J_z}{(\delta-J_z)^3 T^2}\notag \\&-& \frac{45 \zeta(5)}{16\pi^6 \bm{\beta}}  \frac{\delta^4 J_z^2}{(\delta-J_z)^4 T^3}
\left [ 1+ \frac{2}{3} \frac{\delta b^2}{J_z T(\delta-J_z)}\right ]
 .\quad
\label{eq:chizz:regI}
\end{eqnarray}

In the region I the corrections to the longitudinal spin susceptibility are always small and, therefore, the perturbation theory is well justified. We present a more transparent way for derivation of Eq. \eqref{eq:chizz:regI}. At first, one can substitute $1/\Delta$ for $1/\delta$ in the expression \eqref{eq:chizzT0} (with $J_\perp=0$) for the equidistant spectrum. Secondly, we expand $\chi_{zz}$ to the second order in the deviation $\Delta-\delta$. Finally, one can perform averaging with the help of the relation \cite{BGK2}
\begin{equation}
\overline{(\Delta-\delta)^2} = \frac{3\zeta(3)}{2\pi^4\bm{\beta}} \frac{\delta^4}{T^2}, \qquad \delta \ll T   \label{eq:FlucRes}
\end{equation}
and obtain the result \eqref{eq:chizz:regI} (with $b=0$).

In the region II, $\bar{J}_z \gg T \gg \max\{\delta, \bar{J}_z (b/J_z)^2\}$, one can perform an expansion in $x^2 y$ in the right hand side of Eq. \eqref{AverLnXi1} since the condition $1\ll y \ll 1/x^2$ holds. However, the argument of $L$ is typically large and we need to use its asymptotics for $|h|\gg1$ (see Eq.~\eqref{Sassymp}). Then we obtain
\begin{equation}
\overline{\ln \Xi(x,y)} = \frac{y \ln 2}{4\pi^2 \bm{\beta}} \bigl ( 2 + y x^2\bigr ) - \frac{y^3x^4}{48 \pi^2\bm{\beta}}.
\label{eq:Xi:regIII}
\end{equation}
Therefore, the average longitudinal spin susceptibility in the region II ($\bar{J}_z \gg T \gg \max\{\delta, \bar{J}_z (b/J_z)^2\})$ is as follows:
\begin{equation}
\overline{\chi}_{zz} = \frac{1}{2(\delta-J_z)} + \frac{\ln 2}{2 \bm{\beta}\pi^2} \frac{\delta^2}{T(\delta-J_z)^2} - \frac{1}{4\pi^2\bm{\beta}} \frac{\delta^3 b^2}{J_z(\delta-J_z)^3 T^2} .
\label{eq:chizz:regII}
\end{equation}

At zero magnetic field we check that the contribution of  the second order in $L$ to $\overline{\ln \Xi(0,y)}$ is of order of $(y/(\pi^2\bm{\beta}))^2$ (see Appendix \ref{app:2dOrder}). Therefore the perturbation theory in the two-point correlation function of $V$ is justified for $T\gg \bar{J}_z/(\pi^2\bm{\beta})$ only. In this regime the variance of $\chi_{zz}$ is small $\bigl [\overline{(\chi_{zz})^2} - \overline{\chi}_{zz}^2\bigr ]/\overline{\chi}_{zz}^2 \sim \bar{J_z}/(\pi^2 \bm{\beta} T) \ll 1$ (see Appendix \ref{app:2dOrder}). Therefore, at $T\gg \bar{J}_z/(\pi^2\bm{\beta})$ one can expect the normal distribution of $\chi_{zz}$.

Finally, in the region III, $\delta \ll T \ll \bar{J}_z \min\{(b/J_z),(b/J_z)^2\}$, the typical value of $u$ contributing to the integral in the right hand side of Eq. \eqref{AverLnXi1} can be not only of the order unity but also of the order of $x\sqrt{y} \gg 1$. In the latter case, since $y x\gg 1$ one needs to use the asymptotics of $L(h)$ for $|h|\gg1$ (see Eq.~\eqref{Sassymp}). Then we find
\begin{align}
\overline{\ln \Xi(x,y)}  & = \frac{y}{2\pi^2 \bm{\beta}} \Bigl ( \ln \frac{|x| y}{2} + c_2 \Bigr ) -
\int \limits_0^\infty \frac{d u}{\sqrt{\pi}} \, e^{-u^2} L(u\sqrt{2y}) .
\label{eq:Xi:regIII}
\end{align}
We thus obtain the average longitudinal spin susceptibility in the region III ($\delta \ll T \ll  \bar{J}_z \min\{(b/J_z),(b/J_z)^2\}$):
\begin{equation}
\overline{\chi}_{zz} = \frac{1}{2(\delta-J_z)} - \frac{1}{2 \bm{\beta}\pi^2}  \frac{\delta J_z}{(\delta-J_z) b^2} .
\label{eq:chizz:regIII}
\end{equation}
For magnetic fields $b\gg J_z$ the effect of level fluctuations is suppressed and the perturbation theory is justified. At $b \sim  J_z \sqrt{T/\bar{J}_z} \ll J_z$ the result \eqref{eq:chizz:regIII} agrees with the result  \eqref{eq:chizz:regI} whereas at $T \sim \bar{J}_z b/J_z \gg \bar{J}_z$ the corrections due to level fluctuations in \eqref{eq:chizz:regIII} and \eqref{eq:chizz:regII} become of the same order.

\begin{figure}[t]
\includegraphics[width=6cm]{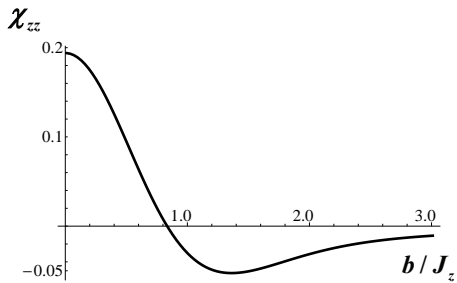}
\caption{Dependence of relative correction $\delta\chi_{zz}=\overline{\chi}_{zz} - 1/[2(\delta-J_z)]$ due to fluctuations on $b/J_z$ (see Eqs. \eqref{eq:chizz:regII} and \eqref{eq:chizz:regIII}). The temperature  $T=\delta J_z/[6(\delta-J_z)]$.}
\label{Fig:chi_zz_on_b}
\end{figure}

Results \eqref{eq:chizz:regII} and \eqref{eq:chizz:regIII} imply non-monotonous behavior of the average longitudinal spin susceptibility with magnetic field $b$  in the temperature range $\bar{J}_z/(\pi^2\bm{\beta}) \ll T \ll \bar{J}_z$ (see Fig. \ref{Fig:chi_zz_on_b}). The susceptibility $\overline{\chi}_{zz}(b)$ as a function of $b$ has a minimum at $b \sim T J_z/\bar{J}_z$. In the region of strong fluctuations $\delta \ll T \ll \bar{J}_z/(\pi^2\bm{\beta})$ we expect similar behavior of the average longitudinal spin susceptibility.

Although the result \eqref{eq:chizz:regIII} is derived for $T\gg \delta$, for $\delta-J_z\ll J_z$ it can be obtained from the following zero-temperature arguments. The difference in the ground state energies for the state with projections $S_z+1$ and $S_z$ of the total spin can be estimated as
\begin{equation}
E_g(S_z+1)-E_g(S_z) = 2(\delta-J_z) S_z -b S_z+\Delta E_{2S_z} .\label{eq:Crit2_0}
\end{equation}
Here $\Delta E_{2S_z}$ is the fluctuation of the energy window in which there are $2S_z$ levels on average. It can be expressed as $\Delta E_{2S_z}=\delta \, \Delta n_{2S_z}$ where $\Delta n_{2S_z}$ is the fluctuation of the number of single-particle levels in the strip with $2S_z$ levels in average. From the random matrix theory it is well known that~\cite{Mehta}
\begin{equation}
\overline{(\Delta n_{2S_z})^2} = \frac{2}{ \pi^2\bm{\beta}} \Bigl [ \ln 2S_z +{\rm \, const}\Bigr ] . \label{eq:Crit2}
\end{equation}
Comparing the energies of the ground states with total spin projections $S_z+1$ and $S_z$, we find from
Eq. \eqref{eq:Crit2_0} that
\begin{equation}
S_z = \frac{1}{2(\delta-J_z)} \Bigl [ b - \delta\, \Delta n_{2S_z}\Bigr ] . \label{eq:CritS3}
\end{equation}
Hence the average longitudinal spin susceptibility can be estimated as
\begin{equation}
\overline{\chi}_{zz} \sim \frac{\partial \overline{S}_z}{\partial b}  =
\frac{1}{2 (\delta-J_z)} \left [ 1 + \frac{\delta^2}{2(\delta-J_z)^2} \frac{d^2 \overline{(\Delta n_z)^2}}{d z^2}\right ] ,
\end{equation}
where $z=2\overline{S}_z\approx b/(\delta-J_z)$. Using Eq.~\eqref{eq:Crit2}, we reproduce the result~\eqref{eq:chizz:regIII}.

\subsubsection{Distribution function for $\chi_{zz}$}

The average longitudinal spin susceptibility is mostly affected by the level fluctuations in the region II ($\bar{J}_z \gg T \gg \bar{J}_z (b/J_z)^2$). The perturbative result \eqref{eq:chizz:regI} loses its validity at $\bar{J}_z/(\pi^2 \bm{\beta}) \gg T \gg \delta$. Such a regime is realized in the close vicinity of the Stoner instability $\delta-J_z\ll  \delta/(\pi^2 \bm{\beta})$. In this case of strong fluctuations it is useful to know the distribution function of $\chi_{zz}$ rather than the average value.

\begin{figure}[t]
\centerline{\includegraphics[width=6cm]{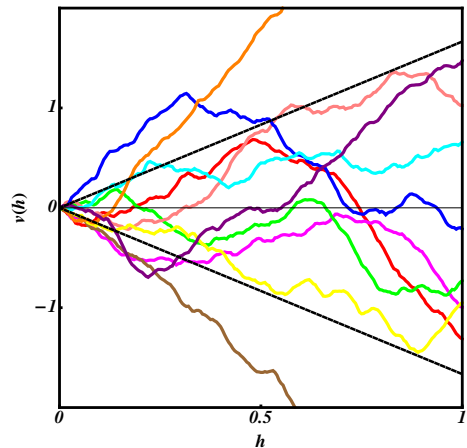}}
\caption{(Color online) Several realizations of the process ${v}(h)$; dashed lines $\pm 2h
\sqrt{\ln 2}$ are guides for the eye.}
\label{Fig:Process}
\end{figure}

In the range of temperatures $\delta \ll T \ll \bar{J}_z$, the integral in the right hand side of Eq. \eqref{eq:Xi:def} is dominated by large values of $|h|$. Then, using the asymptotic expression \eqref{Sassymp}, one can check that for $|h_1|, |h_2|\gg 1$ the two-point correlation function \eqref{corrVV} is homogeneous of degree two:~\cite{KAA}
\begin{equation}
\overline{V(u h_1)V(u h_2)} = u^2 \overline{V(h_1) V(h_2)} .
\label{eq:hom:2deg}
\end{equation}
With the help of Eq. \eqref{eq:hom:2deg}, at zero magnetic field $b=0$ and for $\delta \ll T \ll \bar{J}_z/(\pi^2 \bm{\beta})$, Eqs. \eqref{eq:ZS:Xi} and \eqref{eq:Xi:def} can be simplified to
\begin{equation}
Z_S = \left ( \frac{\delta}{\delta-J_z}\right )^{1/2}\int \limits_{-\infty}^\infty \frac{dh}{\sqrt{\pi}} e^{-h^2 - z v(h)} .
\label{eq:ZS:fluct}
\end{equation}
We remind that the normalization is such that $Z_S = 1$ at $J_z=0$. According to Eq. \eqref{eqZb1} for $J_\perp=b=0$, the grand canonical partition function increases as $J_z$ increases. Hence it follows that $Z_S\geqslant 1$. According to Eq. \eqref{eq:ZS:fluct}, the statistics of the zero field longitudinal spin susceptibility is determined by the single parameter $z = [\beta \bar{J}_z/(\pi^2 \bm{\beta})]^{1/2}$. The  Gaussian random process $v(h)$ has zero mean and is even in $h$, $v(h)=v(-h)$. Its two-point correlation function reads
\begin{align}
\overline{v(h_1)v(h_2)} & =
\frac{1}{2}\sum_{\sigma=\pm} (h_1+\sigma h_2)^2 \ln (h_1+\sigma h_2)^2 \notag \\
& - h_1^2\ln h_1^2 -h_2^2 \ln h_2^2 .
\label{eq:vvh:def}
\end{align}
Hence we find that
\begin{equation}
\overline{\bigl [ v(h+u)-v(h)\bigr ]^2} = - 2 u^2 \ln |u| + O(u^2)=O(u^{2H})
\label{eq:local}
\end{equation}
for any $H=1-\epsilon<1$. Thus the trajectories of $v(h)$ are continuous and its increments are strongly positively correlated (see Fig.~\ref{Fig:Process}). In fact the process $v(h)$ is in many aspects close to the ballistic one $\tilde v(h)=\xi |h|$ with $\xi$ being a Gaussian random variable (recall that $\tilde v(h)$ is the unique process with $H=1$). The process $v(h)$ has arisen before in a seemingly unrelated context.~\cite{BGT}

We are interested in the complementary cumulative distribution function $\mathcal{P}(W)$, i.e. the probability that $\ln Z_S$ exceeds $W$: $\mathcal{P}(W)\equiv \Prob\{\ln Z_S> W\}$. It has the following properties:
$\mathcal{P}(0)=1$, $\mathcal{P}(\infty)=0$ and $\mathcal{P}(W)$ is monotonously decreasing as $W$ increases.
The average moments of $\ln Z_S$ can be conveniently written as $\overline{[\ln Z_S]^k}= k \int_{0}^{\infty}dW W^{k-1} \mathcal{P}(W)$. Although a closed analytical expression for the complementary cumulative distribution function is not known, we bound  $\mathcal{P}(W)$ from above to prove that all moments of $\ln Z_S$ (and consequently all moments of $\chi_{zz}$) are finite for $J_z<\delta$. At first, we split the Gaussian weight $\exp(-h^2)$ in the integral in the right hand side of Eq.~\eqref{eq:ZS:Xi} and obtain ($0<\gamma<1$ is an arbitrary splitting parameter)
\begin{equation}
Z_S \leqslant \frac{2\sqrt{\bar{J}_z}}{\sqrt{\pi \gamma J_z}} \int\limits_{0}^{\infty} dh \, e^{-{(1-\gamma) h^2}/{\gamma}}   \max\limits_{h\geqslant 0} \Bigl \{e^{-h^2-{z v(h)}/{\sqrt{\gamma}}}\Bigr \} .
\label{eq:split}
\end{equation}
The inequality \eqref{eq:split} allows us to reduce the problem of finding an upper bound for $\mathcal{P}(W)$ to  the statistics of the maxima of the Gaussian process $Y_\gamma(h)=-h^2-(z/\sqrt{\gamma})v(h)$ which locally resembles a fractional Brownian motion with a drift. Indeed, from Eq. \eqref{eq:split} we find
\begin{equation}
\mathcal{P}(W) \leqslant \Prob \Bigl \{ \max\limits_{h\geqslant 0} Y_\gamma(h) > W + \frac{1}{2}\ln \frac{(1-\gamma) J_z}{\bar{J}_z} \Bigr \} .
\end{equation}
To give an upper bound for the probability $\Prob\{ \max\limits_{h\geqslant 0} Y_\gamma(h) > w\}$ we employ
the Slepian's inequality. \cite{Slepian} Let us consider an auxiliary Gaussian process $X(h) = -h^2+(2 z\sqrt{\ln 2}/\sqrt{\gamma})B(h^2)$ where $B(h)$ is the standard Brownian motion ($\overline{B(h)}=2h$; the Hurst exponent $H=1/2$). For any interval $\mathcal{T}$ the sample paths $\{X(h), h\in \mathcal{T}\}$ and
$\{Y_\gamma(h), h\in \mathcal{T}\}$ are bounded. The following relations hold:
\begin{gather}
\overline{X(h)}=\overline{Y_\gamma(h)}, \qquad \overline{X^2(h)} =\overline{Y_\gamma^2(h)}, \notag \\
\overline{[X(h_1)-X(h_2)]^2} \geqslant \overline{[Y_\gamma(h_1)-Y_\gamma(h_2)]^2} .
\end{gather}
The first two equalities are trivially satisfied while the last inequality follows from an  easily verifiable inequality
$\overline{[v(1/2+r)-v(1/2-r)]^2}\leqslant 8 r \ln 2$ for $|r|\leqslant 1/2$. Then the processes $Y_\gamma(h)$ and $X(h)$ satisfy the Slepian's inequality:
\begin{equation}
\Prob\{ \max\limits_{h\geqslant 0} Y_\gamma(h) > w\}\leqslant \Prob\{\max\limits_{h\geqslant 0} X(h) > w\}
\end{equation}
for all real $w$.  Using a well-known result for the Brownian motion with a linear drift (see e.g., Ref. [\onlinecite{Asmussen}])
\begin{equation}
\Prob\{\max\limits_{h\geqslant 0} X(h) > w\} = \exp \left ( - \frac{\gamma w}{2 z^2 \ln 2}\right ), \quad w>0 ,
\end{equation}
we find the following upper bound for the complementary cumulative distribution function:
\begin{equation}
\mathcal{P}(W) \leqslant  \exp\left\{-\frac{\gamma}{2z^2\ln 2}\left [W+\frac{1}{2}\ln \frac{(1-\gamma) J_z}{\bar{J}_z}\right ] \right\} .
\label{eq:tail}
\end{equation}
From Eq. \eqref{eq:tail} it follows that for $\bar{J}_z/(\pi^2\bm{\beta})\gg T\gg \delta$ all moments of $\ln Z_S$ (and hence all moments of $\chi_{zz}$) are finite for $J_z<\delta$. Therefore even in the presence of the strong level fluctuations the Stoner instability occurs at $J_z=\delta$ only. For $J_z<\delta$ and for temperatures $T\gg \delta$ the quantum dot is in the paramagnetic state.

For $z\gg 1$ the saddle-point approximation in Eq. \eqref{eq:ZS:Xi} becomes exact and the statistics of $\ln Z_S$ reduces to the statistics of maxima of the process $Y(h)=-h^2-z v(h)$. As it can be seen from rescaling of $h$, the probability that the maximum of $Y(h)$ exceeds $w$ equals the probability that the maximum of $\widetilde Y(s)=v(s)/(1+s^2)$ defined on $s\geqslant 0$ exceeds $\sqrt{w}/z$. From the results of H\"usler and Piterbarg\cite{HuslerPiterbarg} it follows that the large-$w$ tail of $\Prob\{\max\limits_{h\geqslant 0} Y(h) > w\}$ is determined by a small vicinity of the point $s^*=1$ where the variance of $\widetilde Y(s)$ attains its maximum $\ln 2$. Furthermore, should we have a finite limit
\begin{equation}
\lim\limits_{s, t \to s_*}
\frac{\overline{[\widetilde Y(s) - \widetilde Y(t)]^2}}
{K^2(s-t)} >0
\label{eq:Kdef}
\end{equation}
for some function $K(x)$ regularly varying at $0$ with index $\alpha\in(0,1)$, the precise asymptotics would read
\begin{gather}
\Prob\{\max\limits_{h\geqslant 0} Y(h) > W\} \sim {\rm const}(\alpha)\cdot \frac{ (z^2/W)^{-1}}{K^{-1}\bigl (\sqrt{z^2/W}\bigr)} \notag \\
\times \exp \left [-\frac{W}{2z^2\ln 2}\right ] , \quad W/z^2 \gg 1 .
\label{eq:ProbTail}
\end{gather}
Here $K^{-1}(x)$ stands for the functional inverse of $K(x)$. In our case Eq. \eqref{eq:local} translates into $K(x)=x\sqrt{\ln(1/x)}$ which is regularly varying with index $\alpha=1$ [recall that a function $f(x)$ is regular varying at $x=0$ with index $\alpha$ if $\lim\limits_{t\to 0} f(at)/f(t)=a^{\alpha}$ for any $a>0$]. The result of Ref. [\onlinecite{HuslerPiterbarg}] is therefore not directly applicable, but we believe this to be a technicality. In analogy with a similar situation for fractional Brownian motion, we expect the asymptotics \eqref{eq:ProbTail} to hold with only the $W$-independent factor ${\rm const}(\alpha)$ modified. Note that the exponential part can be tracked to be the tail of a normal distribution with variance $\ln 2$ taken at $\sqrt{W}/z$, and that it had been correctly reproduced by our initial estimate. Therefore we find with logarithmic accuracy that the tail of the complementary cumulative distribution function is given by ($W\gg [\ln 2/(\pi^2 \bm{\beta} )] \delta J_z/[T (\delta-J_z)]$)
\begin{align}
\mathcal{P}(W) & \propto \mathcal{P}_{\rm tail}\left ( \frac{\pi^2 \bm{\beta} T (\delta-J_z)W}{\delta J_z}\right ), \notag \\   \mathcal{P}_{\rm tail}(x) & = \frac{\sqrt{\ln x}}{\sqrt{x}}\exp \left ( - \frac{x}{2 \ln 2} \right ).
\label{eq:Prob:exp}
\end{align}
This result is valid in the temperature range $\bar{J}_z/(\pi^2\bm{\beta})\gg T\gg \delta$ and is consistent with the upper bound \eqref{eq:tail}.

\begin{figure}[t]
\centerline{\includegraphics[width=6cm]{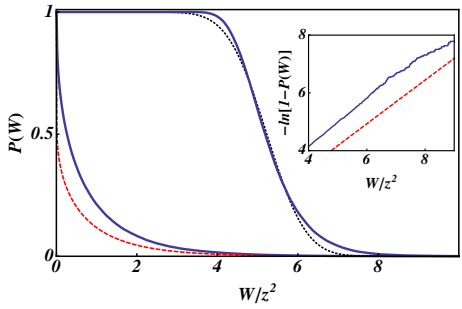}}
\caption{(Color online) The dependence of $\mathcal{P}(W)$ on $W/z^2$ at $T =
3\delta$ computed numerically for $J_z/\delta = 0.94$ ($z \approx 0.5$) (upper solid curve)
and $J_z/\delta = 0.99994$ ($z \approx 16.8$) (lower solid curve). The black dotted
curve is the complementary cumulative distribution function for the normal distribution with mean and variance
as one can find from the lowest order perturbation theory in $V$ for $T =
3\delta$ and $J_z/\delta = 0.94$ (cf. Eqs. \eqref{eq:F2F11} and \eqref{eq:Var2}). The red dashed curve is the complementary cumulative distribution function of the degenerate process $\tilde{v}(h)$ for $T =
3\delta$ and $J_z/\delta = 0.99994$. Inset: Comparison of the tail of $\mathcal{P}(W)$ computed numerically for $J_z/\delta = 0.99994$ ($z \approx 16.8$) and asymptotic result \eqref{eq:Prob:exp}.}
\label{Fig:CCDF}
\end{figure}

To illustrate the result \eqref{eq:Prob:exp} we approximate the Gaussian process $v(h)$ by a degenerate one $\tilde{v}(h) = \xi |h|$, where $\xi$ is the Gaussian random variable with zero mean $\overline{\xi}=0$ and variance $\overline{\xi^2}=4\ln 2$. Substituting the process $\tilde{v}(h)$ for $v(h)$ into the right hand side of Eq. \eqref{eq:ZS:fluct}, we estimate the partition function as $Z_S \simeq \sqrt{\bar{J}_z/J_z} \exp(z^2 \xi^2/4) \Bigl [1 - \erf(z \xi/2) \Bigr ]$. The large values of $Z_S$ correspond to large negative values of $\xi$ such that $\ln Z_S \approx z^2 \xi^2/4$. Therefore, the tail of distribution of $\ln Z_S$ is simple exponential. Hence we find that for $z\gg 1$ the tail of the complementary cumulative distribution function $\mathcal{P}(W)$ is given by Eq.~\eqref{eq:Prob:exp} without the logarithm in the pre-exponent.  As shown in Fig. \ref{Fig:CCDF} the overall behavior of $\mathcal{P}(W)$ for $z\gg 1$ is well enough approximated by the complementary cumulative distribution function for the degenerate process $\tilde{v}(h)$. Also  we mention that the behavior of $\mathcal{P}(W)$ for $z\gg 1$ is very different from its behavior at $z\lesssim 1$. For the later, $\mathcal{P}(W)$ is given by the complementary cumulative distribution function of the normal distribution (see Fig. \ref{Fig:CCDF}).

Equation~\eqref{eq:Prob:exp} implies that the average moments of $\ln Z_S$ scale as $\overline{(\ln Z_S)^k} \sim z^{2k}$ for $z\gg 1$. Hence for $\delta \ll T \ll \bar{J}_z/(\pi^2\bm{\beta})$ the $k$-th moment of the spin susceptibility is given by
\begin{equation}
\overline{\chi_{zz}^k} \propto \left [\frac{\delta^2}{\pi^2 \bm{\beta} (\delta-J_z)^2 T}\right ]^k  , \quad k=1, 2, \dots .
\label{eq:chi:kth}
\end{equation}
The result \eqref{eq:chi:kth} can be obtained from the saddle-point analysis of the integral in the right hand side of Eq. \eqref{eq:ZS:fluct}, i.e., in essence, by Larkin-Imry-Ma type arguments.~\cite{Larkin,ImryMa} The scaling of the average spin susceptibility (Eq. \eqref{eq:chi:kth} with $k=1$) was proposed in Ref. [\onlinecite{KAA}] using arguments of Larkin-Imry-Ma type.

\subsection{The Heisenberg exchange}

For the case of the isotropic exchange, $J_\perp=J_z \equiv J$, the integration over $\mathcal{B}$ in Eq. \eqref{ZS1} becomes trivial. Then for $T\gg \delta$ we obtain \cite{BGK2}
\begin{equation}
Z_S = \left (\frac{\delta}{\delta-J}\right)^{1/2} \frac{e^{\frac{\beta b^2}{4J}}}{\sinh (\beta b/2)}
\tilde{\Xi}\left (\frac{b}{J},\frac{\beta J\delta}{\delta-J}\right) ,
\label{eq:ZS:Xi:iso}
\end{equation}
where
\begin{equation}
\tilde{\Xi}(x,y) = \int\limits_{-\infty}^\infty \frac{dh}{\sqrt{\pi}} \, \sinh(h x \sqrt{y})
e^{-h^2+h \sqrt{y}-V(h\sqrt{y})} .
\label{eq:Xi:def:iso}
\end{equation}
Since in the absence of magnetic field $Z$ grows with increase of $J$ (see Eq. \eqref{eqZb1}), one can check that for the Heisenberg exchange $Z_S\geqslant 1$. The detailed results of the perturbative expansion in $V$ for the longitudinal spin susceptibility can be found in Ref. [\onlinecite{BGK2}]. Similarly to the case of the Ising exchange, the effect of fluctuations is important at $b=0$ and $\delta \ll T\ll J\delta/[\pi^2 \bm{\beta} (\delta-J)]$. In this range  of parameters the typical value of $|h|$ in the integral in the right hand side of Eq. \eqref{eq:Xi:def:iso} is large, $|h| \sim \sqrt{\beta \bar{J}} \gg 1$ where $\bar{J}=\delta J/(\delta-J)$. Then for $b=0$ Eq. \eqref{eqZb1} can be rewritten as
\begin{equation}
Z_S =\frac{2}{\sqrt{\beta J}}\frac{\delta}{\delta-J} \int\limits_{-\infty}^\infty \frac{dh}{\sqrt{\pi}} \, h\, e^{-h^2+h \sqrt{y}-z v(h)} ,
\label{eq:Zs:b0}
\end{equation}
where $z=[\beta \bar{J}/(\pi^2\bm{\beta})]^{1/2}$. Here $\bm{\beta}=1$ which corresponds to the orthogonal Wigner-Dyson ensemble. The complementary cumulative distribution function $\mathcal{P}(W) = \Prob\{\ln Z_S>W\}$ can be estimated in a similar way as in the previous section. Writing 
\begin{gather}
Z_S  \leqslant \frac{2}{\sqrt{\beta J}}\left (\frac{\delta}{\delta-J} \right )\left [2 \int\limits_{0}^{\infty} \frac{dh\, h \sinh (h \sqrt{y})}{\sqrt{\pi \gamma}} e^{-(1-\gamma) h^2/\gamma}\right] \notag \\
 \times  \max\limits_{h\geqslant 0} \Bigl \{e^{-h^2-(z/\sqrt{\gamma})v(h)}\Bigr \} ,
\label{eq:split:iso}
\end{gather}
with arbitrary splitting parameter $\gamma$ ($0<\gamma<1$), we obtain the following upper bound:
\begin{align}
\mathcal{P}(W) \leqslant  a_\gamma   \exp\Biggl\{-\frac{\gamma}{2 z^2 \ln 2} \Biggl
[W+\frac{3}{2}\ln \frac{(1-\gamma)J}{\bar{J}} \Biggr ] \Biggr \} ,
\label{eq:tail:iso}
\end{align}
where $a_\gamma = \exp\{(\pi^2 \bm{\beta} \gamma)/[8 (1-\gamma)\ln 2]\}$.
This upper bound implies that all moments of $\ln Z_S$ (and of $\chi_{zz}$) are finite for $J<\delta$.
At $z \gg 1$ the integral in the right hand side of Eq. \eqref{eq:Xi:def:iso} can be evaluated in the saddle-point approximation, reducing the  statistics of $\ln Z_S$ to the statistics of maxima of the process $Y(h) = - h^2 + h\sqrt{y} - z v(h)$. Then using as in the previous section the results of H\"usler and Piterbarg\cite{HuslerPiterbarg}, we find the tail of the complementary cumulative distribution function at $W \gg \delta J/[T (\delta-J)]$ is given by $\mathcal{P}_{\rm tail}\bigl (\pi^2 \bm{\beta} T (\delta-J)W/(\delta J)\bigr )$ (see Eq. \eqref{eq:Prob:exp}). We note that for this tail the drift term $h \sqrt{y}$ in the process $Y(h)$ is not important.

The typical value of $h$ contributing to the integral in the right hand side of Eq. \eqref{eq:Zs:b0} is $\sqrt{y}/2$.
Then for $z\gg 1$ we find, with logarithmic accuracy,  $\ln Z_s - y/4 = (z \sqrt{y}/2) v(1)$. Hence for $\delta \ll T \ll \bar{J}/(\pi^2\bm{\beta})$ the average $k$-th moment of the longitudinal spin susceptibility can be estimated as
\begin{equation}
\overline{\left (\chi_{zz} - \chi^{(0)}_{zz}\right )^k} \propto \left (\frac{\delta^2}{\sqrt{\pi^2 \bm{\beta}} T(\delta-J)^2}\right )^k ,
\label{eq:chi:kth:iso}
\end{equation}
where $\chi^{(0)}_{zz} = \delta^2/[12 T (\delta-J)^2]$ is the spin susceptibility in the absence of level fluctuations. We note that for $\delta \ll T \ll \bar{J}/(\pi^2\bm{\beta})$ the scaling of the average longitudinal spin susceptibility similar to Eq. \eqref{eq:chi:kth:iso} with $k=1$ was derived in Ref. [\onlinecite{KAA}].

\section{Transverse spin susceptibility \label{sec:trss}}

The transverse spin susceptibility is defined as follows (see, e.g., Ref. [\onlinecite{Mahan}])
\begin{equation}
\chi_{\perp}(\omega)=\frac{i}{Z}\int\limits_{0}^\infty dt e^{i(\omega+i0^+) t} \Tr\Bigl ([\hat{S}_+(t),\hat{S}_-(0)]e^{-\beta H}\Bigr ),
\label{eq:chi_perp_def}
\end{equation}
where $\hat{S}_\pm=\hat{S}_x\pm i\hat{S}_y$.  Since, in contrast with $\hat{S}_z$, the operators $\hat{S}_x$, $\hat{S}_y$ of the total spin do not commute with the Hamiltonian $H$ (for $J_z\neq 0$), the transverse spin susceptibility can acquire non-trivial frequency dependence.

In order to find the dynamic transverse spin susceptibility \eqref{eq:chi_perp_def} we use the Heisenberg equations of motion for the spin operators: $d\hat{\bm{S}}/dt = i [H, \bm{S}]$. Since the operator $S_z$ commutes with the Hamiltonian, it has no dynamics, $\hat{S}_z(t)=\hat{S}_z$. For the other components of the total spin we find
\begin{align}
\hat{S}^\pm(t) &=e^{\mp 2i (J_\perp-J_z) \hat{S}_z t} \hat{S}^\pm(0)e^{-i(J_\perp-J_z)t\pm i b t} \notag \\
&\equiv\hat{S}^\pm(0) e^{\mp 2i (J_\perp-J_z)  \hat{S}_z t} e^{i(J_\perp-J_z)t\pm i b t} .
\label{eq:Spmt}
\end{align}
Using expressions \eqref{eq:Spmt}, we integrate over time in Eq. \eqref{eq:chi_perp_def} and obtain the following operator expression for the transverse spin susceptibility:
\begin{align}
\chi_{\perp}(\omega)= \frac{1}{Z}
\sum_{\sigma=\pm} \Tr \frac{\Bigl (\sigma \bigl [\bm{\hat{S}}^2-\hat{S}_z^2\bigr ]-\hat{S}_z\Bigr )e^{-\beta H} }{\omega+b+(J_\perp-J_z)(2\hat{S}_z+\sigma)+i0^+} .
\label{eq:chi_perp_2}
\end{align}
Since operators $\hat{S}_z$ and $\bm{\hat{S}}^2$ commute with $H$, one easily evaluates the trace in Eq. \eqref{eq:chi_perp_2} with the help of Eq. \eqref{eqZb1}. Thus we derive the exact result for the dynamic transverse spin susceptibility:
\begin{align}
\chi_\perp(\omega) & =\frac{1}{Z} \sum\limits_{n_\uparrow,n_\downarrow} Z_{n_\uparrow}Z_{n_\downarrow}
e^{-\beta E_c(n-N_0)^2+\beta J_\perp m(m+1)+\beta\mu n} \notag \\
& \times
\sgn \left (2m+1\right )  \sum\limits_{l=-|m+1/2|+1/2}^{|m+1/2|-1/2}  e^{\beta(J_z-J_\perp)l^2 - \beta b l}
\notag \\
& \times
\sum\limits_{\sigma=\pm}
\frac{\sigma \bigl [ m(m+1)-l^2\bigr ]-l}{\omega+b+(J_\perp-J_z)(2l+\sigma)+i0^+} .
\label{eqZb2}
\end{align}
In what follows we will be interested in the imaginary part of $\chi_\perp(\omega)$. The real part can be restored from the Kramers-Kronig relations. Using Eq.~\eqref{eqZb2}, the imaginary part of the dynamic transverse spin susceptibility can be written as
\begin{align}
\Imag \chi_{\perp}(\omega) & = - \frac{\pi}{Z}\sum\limits_{n\in\mathbb{Z}}  \sum\limits_{\sigma=\pm} \delta\Bigl (\omega +b+(2n-\sigma)(J_z-J_\perp)\Bigr )\notag \\
& \times \Bigl ( n + \sigma T \frac{\partial}{\partial J_\perp}\Bigr ) Z(n) .
\label{eq:DSS1}
\end{align}
Here we introduce the Fourier transform of the partition function $Z(b+i\lambda T)$ in the complex magnetic field $b+i\lambda T$:
\begin{equation}
Z(n)=  \int\limits_{-\pi}^\pi \frac{d\lambda}{2\pi} \, e^{-i\lambda n} Z(b+i\lambda T) .
\label{eq:DSS10}
\end{equation}
As it follows from Eq. \eqref{eq:DSS1}, the imaginary part of the transverse spin susceptibility obeys the sum rule
\begin{equation}
\int \limits_{-\infty}^\infty \frac{d\omega}{2\pi} \Imag \chi_\perp(\omega) = M ,
\label{eq:sumrule_chi_perp}
\end{equation}
where the magnetization $M = -\langle \hat{S}_z\rangle =T \partial \ln Z/\partial b$. Since at $b=0$ the function $Z(n)$ is even, the imaginary part of the zero-field transverse spin susceptibility is odd in frequency: $\Imag \chi_\perp(-\omega) = -\Imag \chi_\perp(\omega)$, so the sum rule \eqref{eq:sumrule_chi_perp} is trivially satisfied.

We mention that in the case of an isotropic exchange, $J_z=J_\perp$, Eq. \eqref{eq:DSS1} becomes trivial, $\Imag \chi_{\perp}(\omega) = 2\pi M \delta(\omega-b)$. In this case the behavior of the transverse spin susceptibility is fully determined by the behavior of the magnetization $M$. Therefore, in what follows we shall not discuss the transverse spin susceptibility for the isotropic exchange.

\subsection{Equidistant single-particle spectrum}

At first we consider the case of an equidistant single-particle spectrum and, therefore, neglect effects related to the level fluctuations. As it was discussed in Sec. \ref{sec:longsusc}, for $\delta\ll T$ the partition function can be factorized in accordance with Eq. \eqref{ZCZS}. Since the factor $Z_C$ does not depend on the magnetic field, it does not influence the results for $\chi_\perp(\omega)$ and we omit it below in this section. It implies that $Z_S$, $Z_S(n)$, and $Z_S(b+i\lambda T)$ should be substituted for $Z$, $Z(n)$, and $Z(b+i\lambda T)$ in Eqs. \eqref{eq:DSS1} - \eqref{eq:DSS10}, respectively. Using Eq.\eqref{EQm2}  for the equidistant single-particle spectrum we can rewrite $Z_S(n)$ in the following way:
\begin{gather}
Z_S(n)=\sqrt{\frac{\beta \delta}{\pi}}\int_{-\pi}^{\pi} \frac{d\lambda}{2\pi}e^{-i\lambda n}\sum_m e^{-\beta (\delta-J_\perp) m^2+\beta J_\perp m}\notag \\
\times \sgn \left (2m+1\right )  \sum\limits_{l=-|m+1/2|+1/2}^{|m+1/2|-1/2}   e^{\beta(J_z-J_\perp)l^2-\beta bl -i\lambda l}.
\label{eq:Zn:2}
\end{gather}
Next, performing integration over $\lambda$, we obtain the following result:
\begin{align}
Z_S(n)&=\sqrt{\frac{\beta \delta}{\pi}} e^{\beta(J_z-J_\perp)n^2+\beta b n} \Biggl [ \sum_{m=|n|} e^{-\beta (\delta-J_\perp) m^2+\beta J_\perp m}  \notag \\
& -\sum_{m=|n|+1} e^{-\beta (\delta-J_\perp) m^2-\beta J_\perp m} \Biggr ]
.
\label{Im_General}
\end{align}
In the case $\beta (\delta-J_\perp) |n| \ll 1$, applying the Euler-Maclaurin formula to estimate the sums over $m$, we find
\begin{align}
Z_S(n) & = \frac12 \left (\frac{\delta}{\delta-J_\perp}\right )^{1/2} e^{\frac{\beta J_\perp^2}{4(\delta-J_\perp)}}e^{\beta(J_z-J_\perp)n^2+\beta b n} \notag \\
&  \times \sum\limits_{s=\pm} \erf \left (\sqrt{\beta(\delta-J_\perp)} \left(s |n| + \frac{J_\perp}{2(\delta-J_\perp)}\right )\right ) \notag \\
& +  \sqrt{\frac{\beta \delta}{\pi}} e^{-\beta(\delta-J_z)n^2+\beta b n} \cosh (\beta J_\perp |n|)
.
\label{eq:ZSn:small}
\end{align}
In the opposite case $\beta (\delta-J_\perp) |n| \gg 1$, the term with $m=|n|$ in the right hand side of Eq. \eqref{Im_General} provides the main contribution. Then we obtain
\begin{equation}
Z_S(n) = \sqrt{\frac{\beta \delta}{\pi}} e^{-\beta(\delta-J_z)n^2+\beta J_\perp|n| + \beta b n}  .
\label{eq:ZSn:large}
\end{equation}
We note that for $J_\perp=0$, both expressions \eqref{eq:ZSn:small} and \eqref{eq:ZSn:large} coincide and are valid, in fact, for arbitrary $n$.

According to Eq. \eqref{eq:DSS1}, $\Imag \chi_{\perp}(\omega)$ is represented as the sum of delta-peaks. Since their positions are independent of the realization of single-particle levels, the delta-peaks survive averaging of $\Imag \chi_{\perp}(\omega)$ over level fluctuations. Therefore, in order to discuss the frequency dependence of the transverse spin susceptibility in a form of a smooth curve, we assume some natural broadening $\Gamma \gg |J_z-J_\perp|$ for these delta-peaks. Then the sum over $n$ in Eq. \eqref{eq:DSS1} can be replaced by an integral and we obtain
\begin{align}
\Imag \chi_{\perp}(\omega)  & = -\frac{\pi}{2|J_z-J_\perp|Z_S} \sum\limits_{\sigma=\pm}
\notag \\
& \times \Bigl ( n + \sigma T \frac{\partial}{\partial J_\perp}\Bigr ) Z_S(n) \Biggl |_{n = -\varpi+\sigma/2},
\label{eq:DSS1_Int}
\end{align}
where $\varpi = (\omega +b)/[2(J_z-J_\perp)]$.

In the limit of large frequencies or large magnetic fields,  $\beta (\delta-J_\perp) |\varpi|\gg 1$, the imaginary part of the transverse spin susceptibility is exponentially small:
\begin{gather}
\Imag \chi_{\perp}(\omega)   =  \frac{\varpi \sqrt{\pi\beta\delta}}{|J_z-J_\perp|Z_S}  \exp \Biggl [ -\beta(\delta-J_z) |\varpi|(|\varpi|+1) \notag \\
  +\beta J_\perp |\varpi|- \beta b \varpi \Biggr ] .
  \label{eq:ImagChi:Tail}
\end{gather}

In the absence of magnetic field, $b=0$, $\Imag \chi_{\perp}$ is an odd function of the frequency $\omega$. For $\omega \to 0$ the imaginary part of the transverse spin susceptibility has linear behavior:
\begin{align}
\Imag  \chi_{\perp}(\omega)  & =  \frac{\omega \sqrt{\pi\beta \delta}}{2|J_z-J_\perp|(\delta-J_\perp)Z_S} \Biggl [
\frac{2\delta-J_\perp}{2(\delta-J_\perp)}+
\notag \\
& + \frac{\sqrt{\pi}}{2\sqrt{\beta(\delta-J_\perp)}}  \mathcal{G}\left (\frac{\beta J^2_\perp}{4(\delta-J_\perp)}\right )\Biggr ]
\end{align}
where
\begin{equation}
\mathcal{G}(x) = (1+2x) e^{x} \erf(\sqrt{x}) .
\end{equation}

\begin{figure}[t]
\centerline{\includegraphics[width=6cm]{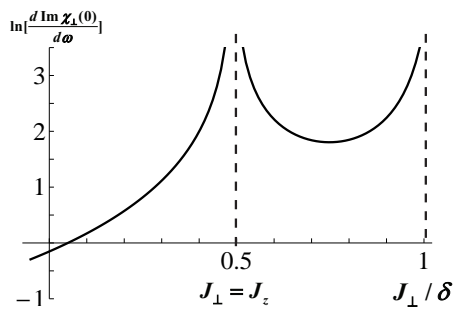}}
\caption{Dependence of the slope $\frac{d\Imag  \chi_{\perp}(0)}{d\omega}$ of the traverse spin susceptibility on $J_\perp/\delta$ for $J_z=\delta/2$.}
\label{Fig:ImChiSlope}
\end{figure}

The slope of $\Imag  \chi_{\perp}(\omega)$ at $\omega=0$ has different behaviors for $J_\perp<J_z$  and for $J_\perp > J_z$. In the interval $0\leqslant J_\perp\leqslant J_z$  the slope grows monotonously with the increase of $J_\perp$ and diverges at $J_\perp=J_z$. In the range $J_z<J_\perp<\delta$ the slope has a minimum (see Fig. \ref{Fig:ImChiSlope}).  The imaginary part of the zero field transverse spin susceptibility has two extrema (a minimum at a negative frequency and a maximum at a positive frequency). In the case $\delta-J_z, J_\perp \ll \delta$ and $\delta\ll T\ll \delta^2/(\delta-J_z)$ the positions of the extrema can be estimated as
\begin{gather}
\omega_{\rm ext} \approx \pm \frac{2(J_z-J_\perp)}{\sqrt{2\beta(\delta-J_z)}} \Biggl [
\left ( 1+\frac{\beta J_\perp^2}{8(\delta-J_z)}\right )^{1/2} \notag \\
+\left(\frac{\beta J_\perp^2}{8(\delta-J_z)}\right )^{1/2}
\Biggr ]  .
\label{eq:ext:gen}
\end{gather}

\begin{figure}
\includegraphics[width=6cm]{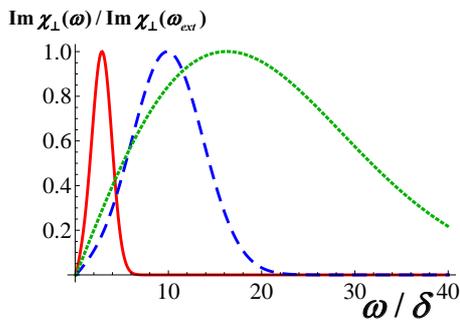}
\caption{(Color online) Dependence of $\Imag \chi_{\perp}(\omega)$ on $\omega$ for $J_z=0.98\delta$ and several values of  $J_\perp$: $J_\perp=0.92\delta$ (red solid line), $J_\perp=0.75\delta$ (blue dashed line) and $J_\perp=0$(green dotted line). The curves shrink to $\omega=0$ as one moves closer to the isotropic case.}
\label{fig:DSSplots}
\end{figure}

The behavior of $\chi_\perp(\omega)$ as a function of frequency is shown in Fig. \ref{fig:DSSplots}. In the presence of a magnetic field $\Imag  \chi_{\perp}(\omega)$ is shifted along the frequency axis and becomes asymmetric (see Fig.  \ref{fig:DSSplots}).

It is worthwhile to discuss the case of the Ising exchange ($J_\perp=0$) in more detail. In the regime of small frequencies and magnetic fields, $|\omega|, |b| \ll T J_z/\delta$, the imaginary part of the dynamic spin susceptibility reads
\begin{align}
\Imag \chi_\perp(\omega) = \frac{\omega \sqrt{\pi \beta (\delta-J_z)}}{2 J_z \delta} \exp \left \{-\frac{\beta[(\delta-J_z)\omega+\delta b]^2}{4J_z^2(\delta-J_z)}\right \} .
 \label{eqnIsing2S}
\end{align}
Although $\Imag \chi_\perp(\omega)$ is asymmetric in the presence of magnetic field, it still vanishes at zero frequency, $\Imag \chi_\perp(\omega=0)=0$. In the opposite limit $|\omega|, |b| \gg T J_z/\delta$, from Eq. \eqref{eq:ImagChi:Tail} we find
\begin{align}
\Imag \chi_\perp(\omega) & = \frac{(\omega +b) \sqrt{\pi \beta (\delta-J_z)}}{2 J_z^2} \exp\left \{-\frac{\beta(\delta-J_z)}{2J_z}|\omega+b| \right \}\notag \\
& \times\exp \left \{-\frac{\beta[(\delta-J_z)\omega+\delta b]^2}{4J_z^2(\delta-J_z)}\right \} .
\label{eqnIsing2L}
\end{align}

In the case $b=0$, our results~\eqref{eqnIsing2S} and \eqref{eqnIsing2L} coincide with the small and large frequency asymptotics of the result obtained in Ref.~[\onlinecite{Boaz}]. The presence of a magnetic field leads to a shift of the extrema of the imaginary part of the dynamic spin susceptibility according to
\begin{gather}
\omega_{\rm ext} \approx\pm \sqrt{ \frac{2J_z^2}{\beta(\delta-J_z)}}  \Biggl \{\left (1+\frac{\beta b^2}{8(\delta-J_z)}\right )^{1/2}
\notag \\
\mp \frac{b\sqrt\beta}{8\sqrt{(\delta-J_z)}} \Biggr \} .
\label{eq:max:Ising}
\end{gather}

\subsection{The effect of level fluctuations (Ising case)}

The above results for the dynamic spin susceptibility have been obtained without taking the level fluctuations into account. Below we consider how the level fluctuations affect the dynamic spin susceptibility in the case of the Ising exchange. As we shall demonstrate, the effect of level fluctuations on $\Imag \chi_\perp(\omega)$ is small in most cases. Since the effect of level fluctuations is suppressed by the magnetic field, below we consider only the case $b=0$.

We start from a generalization of Eq. \eqref{EQm2} to an arbitrary spectrum (see Appendix \ref{App:ZnZn}):
\begin{align}
Z_{n_\uparrow} Z_{n_\downarrow} \approx & \sqrt{\frac{\beta \delta}{4\pi}}
e^{-\beta\mu_n n - 2 \beta \Omega_0 (\mu_n)} \notag \\
& \times \int\limits_{-\infty}^\infty \frac{d\theta}{\pi} e^{-2 m i \theta} e^{-\frac{\theta^2}{\beta\delta} - V(i\theta)} .
\label{EQm2:gen}
\end{align}
With the help of Eqs. \eqref{eqZb2}, \eqref{eq:DSS1_Int}, and \eqref{EQm2:gen}
we rewrite the imaginary part of the dynamic spin susceptibility as follows:
\begin{gather}
\Imag \chi_\perp(\omega)  =  - \frac{\sqrt{\pi\beta(\delta-J_z)}}{2J_z} \sum\limits_{\sigma=\pm} e^{\beta J_z n^2}    \Biggl [ \sum\limits_{m=|n|+1}^\infty 2 \sigma  m \notag\\
\times e^{-\beta \delta m^2} F_\chi\left (m,\beta\delta,\frac{\beta \delta J_z}{\delta-J_z}\right )
+ (n+\sigma |n|) e^{-\beta \delta n^2} \notag \\
\times F_\chi\left (|n|,\beta\delta,\frac{\beta \delta J_z}{\delta-J_z}\right )
 \Biggr ] \Biggl |_{n=(\sigma J_z-\omega)/(2J_z)} .
\end{gather}
Here the random function
\begin{align}
F_\chi(m,x,y) & = \frac{\int\limits_{-\infty}^\infty d\theta\,
e^{-\theta^2-V\left (x m+i\theta\sqrt{x}\right )}}{\int\limits_{-\infty}^\infty dh\,e^{-h^2-V\left (h\sqrt{y}\right )}}\label{eq:def:Fchi}
\end{align}
is equal to unity in the absence of level fluctuations (for $V=0$).

Expanding the right hand side of Eq. \eqref{eq:def:Fchi} to the second order in $V$ we find
\begin{align}
\overline{F_\chi(m)} = & \int\limits_{-\infty}^\infty \frac{dh_1d h_2}{\pi} e^{-h_1^2-h_2^2}  \Biggl \{
1
+\frac{1}{2} L\left (2x m+2 i h_1\sqrt{x}\right ) \notag \\
 & -  2 L\left (x m+i h_1\sqrt{x}+h_2\sqrt{y}\right )
- \frac{1}{2} L\left (2h_1\sqrt{y}\right )
\notag \\ & + 2L \left (h_1\sqrt{y}+ h_2\sqrt{y}\right )
\Biggr \} .
\label{eq:Fchi:exp}
\end{align}

In the high temperature regime, $T\gg \delta J_z/(\delta-J_z)$, and for $|m|\ll T/\delta$, all three integrals in the right hand side of Eq. \eqref{eq:Fchi:exp} are of the same order. Using the asymptotic expression \eqref{Sassymp} for the function $L(h)$ at $|h|\ll 1$, we obtain the following result for the imaginary part of the average dynamic spin susceptibility at low frequencies $\delta |\omega|/(2J_z)\ll T$ and high temperatures $T\gg \delta J_z/(\delta-J_z)$:
\begin{align}
\frac{\Imag \overline{\chi_\perp(\omega)}}{\Imag \chi^{(0)}_\perp(\omega)}  =
1 & + \frac{3\zeta(3)\delta^2}{16\pi^4\bm{\beta} T^2} \Biggl [ -\frac{\delta^2}{(\delta-J_z)^2}\notag \\
& -\frac{\delta^2\omega^2}{TJ_z^2(\delta-J_z)}+ \frac{\delta^2\omega^4}{4T^2J_z^4}
\Biggr ] .
\label{eq:Ising:Fluc1}
\end{align}

Here $\Imag \chi^{(0)}_\perp(\omega)$ is given by Eq. \eqref{eqnIsing2S} with $b=0$. We mention that Eq. \eqref{eq:Ising:Fluc1} can be obtained from Eq. \eqref{eqnIsing2S} if one substitutes $1/\Delta$ for $1/\delta$ and performs averaging with the help of Eq. \eqref{eq:FlucRes}. In the regime of low frequencies and high temperatures the effect of level fluctuations is small.

In the case of high frequencies  and high temperatures, $\delta |\omega|/(2J_z)\gg T \gg \delta J_z/(\delta-J_z)$, the first and second lines in the right hand side of Eq. \eqref{eq:Fchi:exp} provide the main contribution. Then with the help of the asymptotic expression \eqref{Sassymp} for $L(h)$ at $|h|\gg 1$ we find that for $|\omega|/(2J_z)\gg T/\delta \gg J_z/(\delta-J_z)$ the imaginary part of the average dynamic spin susceptibility can be written as
\begin{gather}
\frac{\Imag \overline{\chi_\perp(\omega)}}{\Imag \chi^{(0)}_\perp(\omega)} =  1 + \frac{\ln 2}{2\pi^2\bm{\beta}}
\frac{\omega^2\delta^2}{T^2J_z^2} .
\label{eq:Ising:Fluc2}
\end{gather}
Here $\Imag \chi^{(0)}_\perp(\omega)$ is given by Eq. \eqref{eqnIsing2L} with $b=0$. We note that the result \eqref{eq:Ising:Fluc2} is valid provided $[\omega\delta/(T J_z)]^2\ll \pi^2\bm{\beta}$ so that the perturbation theory in $V$ is justified. We emphasize that although the result Eq. \eqref{eq:Ising:Fluc2} is valid at high temperatures $T \gg \delta J_z/(\delta-J_z)$, it cannot be obtained from Eq. \eqref{eqnIsing2L} by a substitution of $1/\Delta$ for $1/\delta$ and averaging with the help of Eq. \eqref{eq:FlucRes}.

In the case of low temperatures $T\ll \delta J_z/(\delta-J_z)$, the $m$-independent contributions in the right hand side of Eq. \eqref{eq:Fchi:exp} vanish in the leading order. Using the asymptotic result for $L(h)$ at $|h|\gg 1$ (see Eq. \eqref{Sassymp}) we obtain

\begin{equation}
\overline{F_\chi(m)}  = 1 -\frac{x}{\pi^2\bm{\beta}} \begin{cases}
(x m^2-\frac12) \ln y, &  x|m|\ll 1, \\
(x m^2-\frac12) \ln \frac{y}{x^2m^2}, & 1\ll x|m| \ll \sqrt{y}, \\
\frac{y}{2x} \ln \frac{x^2 m^2}{y}, & \sqrt{y} \ll x|m| .
\end{cases}
\label{eq:Fchi:asympt}
\end{equation}
Hence we find the following result for the imaginary part of the average dynamical spin susceptibility at low frequencies,  $|\omega|/J_z^2\ll T/\delta\ll J_z/(\delta-J_z)$:
\begin{gather}
\frac{\Imag \overline{\chi_\perp(\omega)}}{\Imag \chi^{(0)}_\perp(\omega)} = 1 - \frac{\delta}{\bm{\beta} \pi^2 T} \left ( \frac{\delta \omega^2}{4 T J_z^2} + \frac{1}{2} \right )
\ln \frac{\delta J_z}{(\delta-J_z) T} .
\label{eq:Ising:Fluc3}
\end{gather}
Here $\Imag \chi^{(0)}_\perp(\omega)$ is given by Eq. \eqref{eqnIsing2S} with $b=0$.
In the temperature range $|\omega|/J_z^2\ll T/\delta\ll J_z/(\delta-J_z)$ the effect of  level fluctuations is suppressed by an additional small factor $\delta/T\ll 1$. Thus we expect that the perturbation theory is valid
even for $T \ll \delta J_z/[\pi^2\bm{\beta}(\delta-J_z)]$.

In the high frequency regime,  $1 \ll (\omega\delta/(J_zT))^2$, and at low temperatures $T\ll \delta J_z/(\delta-J_z)$ we obtain from Eq. \eqref{eq:Fchi:asympt}  the following result for the average dynamical spin susceptibility:
\begin{gather}
\frac{\Imag \overline{\chi_\perp(\omega)}}{\Imag \chi^{(0)}_\perp(\omega)} = 1 + \frac{1}{2\pi^2 \bm{\beta}}
\min\left \{ \frac{\omega^2 \delta^2}{2J_z^2 T^2}, \frac{\delta J_z}{T(\delta-J_z)} \right \} \notag \\
\times \ln \max\left \{ \frac{\omega^2 \delta(\delta-J_z)}{4J_z^3 T}, \frac{4J_z^3 T}{\omega^2 \delta(\delta-J_z)} \right \} .
\label{eq:Ising:Fluc4}
\end{gather}
Here $\Imag \chi^{(0)}_\perp(\omega)$ is given by Eq. \eqref{eqnIsing2L} for $b=0$.  The perturbation theory is justified for $\max \left \{ [\omega\delta/(J_z T)]^2, \delta J_z/[T(\delta-J_z)] \right \} \ll \pi^2\bm{\beta}$. We remind that the maximum of $\Imag \chi^{(0)}_\perp(\omega)$ is close to the frequency $\omega_{\rm ext} \approx \sqrt{2J_z^2 T/(\delta-J_z)}$. Then, as it follows from Eq. \eqref{eq:Ising:Fluc4}, the fluctuations yield an enhancement of the maximal value of the average dynamical spin susceptibility of the relative order $[(\delta J_z/(\pi^2 \bm{\beta}T(\delta-J_z))]$. Due to fluctuations there is a small shift of the maximum towards zero frequency, $\delta \omega_{\rm ext}/\omega_{\rm ext} \sim - \delta^2/(\pi^2 \bm{\beta} T^2)$.

Since $Z_S\leqslant 1$, we can bound the function $F_\chi(m)$ from above as 
\begin{equation}
F_\chi(m) \leqslant \left ( \frac{\delta}{\delta - J_z}\right )^{1/2} \int\limits_{-\infty}^\infty d\theta\,
e^{-\theta^2-V\left (x m+i\theta\sqrt{x}\right )} .
\label{eq:F:bound}
\end{equation}
Therefore $F_\chi(m)$ remains finite for $J_z<\delta$. Thus, in spite of the level fluctuations, the Stoner instability in $\Imag \chi_\perp(\omega)$ emerges only at $J_z=\delta$.

According Eq. \eqref{eq:F:bound}, averaging over level fluctuations keeps $\Imag \chi_\perp(\omega)$ finite. However, the form of the curve can be changed drastically in the regime of strong fluctuations. To estimate  $\Imag \chi_\perp(\omega)$ at $\delta \ll T \ll \delta J_z/[\pi^2\bm{\beta}(\delta-J_z)]$ we substitute the degenerate process $\tilde{v}(h)$ for $V(h)$ into Eq. \eqref{eq:def:Fchi}. Then a straightforward calculation yields
\begin{equation}
\overline{F_\chi(m)}= \frac{e^{ \beta (\delta-J_z) m^2}}{\sqrt{8z^2\ln 2}}\exp \left[ - \frac{\beta (\delta-J_z) m^2}{2 z^2\ln 2} \right ]
\end{equation}
for $\beta (\delta - J_z) m^2 \gg 1$. We recall that $z^2 = \delta^2/[\pi^2 \bm{\beta} T (\delta-J_z)]$.
This result implies that $\Imag \overline{\chi_\perp(\omega)}$ has a minimum and a maximum at frequencies
\begin{equation}
\omega_{\rm ext} = \pm \frac{2 \sqrt{\ln 2}}{\sqrt{\pi^2 \bm{\beta}}} \frac{\delta^2}{\delta-J_z} .
\end{equation}
Due to strong fluctuations of the single-particle levels the frequency of the extremum shifts towards higher frequencies (in comparison with the corresponding result without fluctuations) and becomes temperature independent. The fluctuations do not affect considerably the values of $\Imag  \overline{\chi_\perp(\omega)}$ at the extrema. Therefore the slope at $\omega=0$ becomes smaller, $
\Imag  \overline{\chi_\perp(\omega)}/\Imag  {\chi^{(0)}_\perp(\omega)} \propto 1/z \ll 1$.


\section{Conclusions \label{sec:dc}}

In this paper we have addressed the spin fluctuations and dynamics in quantum dots and ferromagnetic nanoparticles. Within the framework of the model Hamiltonian which is an extension of the universal Hamiltonian to the case of uniaxial anisotropic exchange interaction, we have derived exact analytic expressions for the static longitudinal and dynamic transverse spin susceptibilities for arbitrary single-particle spectrum.

For the equidistant single-particle levels we analyzed the temperature and magnetic field dependence of $\chi_{zz}$. For $J_\perp\neq 0$ the zero-field longitudinal spin susceptibility has temperature dependence of type $1/T$ (Curie-like) or $1/\sqrt{T}$.  This indicates that the destruction of the mesoscopic Stoner instability by uniaxial anisotropy is not abrupt. The magnetic field suppresses the temperature dependence of $\chi_{zz}$ making spins aligned along the field.

For the case of the Ising exchange interaction we study the effect of single-particle level fluctuations on $\chi_{zz}$ in detail. The temperature dependence of $\chi_{zz}$ appears only due to level fluctuations. We showed that at low temperatures and for $\delta-J_z\ll\delta$ (where fluctuations are strong) the statistical properties of the longitudinal spin susceptibility are determined by the statistics of the extrema of a Gaussian process with a drift. This random process resembles locally a fractional Brownian motion. We rigorously prove that in this regime of strong fluctuations all moments of zero-field static longitudinal spin susceptibility $\chi_{zz}$ are finite for $J_z<\delta$ and temperatures $T \gg \delta$. This means that the Stoner instability is not shifted by the level fluctuations away from its average position at $J_z = \delta$. Also, our results imply that randomness in the single-particle levels does not lead to a transition at finite $T\gg\delta$ between a paramagnetic and a ferromagnetic phase. We expect that these conclusions hold also for temperatures $T \lesssim \delta$. However, we cannot argue it within our approach; a separate (perhaps numerical) analysis is needed. We found that the magnetic field suppresses the effect of level fluctuations on the average longitudinal spin susceptibility. Interestingly, the dependence of $\overline{\chi}_{zz}$ on $b$ is non-monotonous with a minimum. We extended the analysis of the effect of strong level fluctuations to the case of Heisenberg exchange. We demonstrated that in this case the very same conclusions as for the Ising exchange hold.

For equidistant single-particles levels we computed the temperature and magnetic field dependence of the imaginary part of the transverse spin susceptibility $\Imag \chi_\perp(\omega)$. We found that it always has a maximum and a minimum whose positions tend to zero frequency with the decrease of anisotropy. The height of the maximum and the depth of the minimum increase with the decrease of anisotropy.

For the Ising exchange we took into account the effect of single-particle level fluctuations on $\Imag \chi_\perp(\omega)$. We argued that all moments of the dynamic transverse spin susceptibility $\chi_\perp(\omega)$ do not diverge for  $J_z<\delta$. We found that at $\delta-J_z\ll \delta$ the positions of the extrema of $\Imag \overline{\chi}_\perp(\omega)$ have a $\sqrt{T}$-type dependence at high temperatures and become independent of $T$ at low temperatures (in the regime of strong level fluctuations). Interestingly, the level fluctuations do not change the minimal and maximal values of $\Imag \overline{\chi}_\perp(\omega)$ significantly.

Our results, in principle, can be checked in quantum dots and nanoparticles made of materials close to the Stoner instability, such as Co impurities in a Pd or Pt host, Fe or Mn dissolved in various transition-metal alloys, Ni impurities in a Pd host, and Co in Fe grains, as well as nearly ferromagnetic rare-earth materials.~\cite{Exp1} However, to test our most interesting results on spin susceptibility in the regime of strong level fluctuations one needs to explore the regime $(\delta-J_z)/\delta\ll 1/(\pi^2\bm{\beta})$. The closest material to the Stoner instability we are aware of, YFe$_2$Zn$_{20}$, \cite{Exp2} has the exchange interaction $J \approx 0.94 \delta$ which is near the border of the regime with strong level fluctuations at low temperatures.


\begin{acknowledgements}

We acknowledge useful discussions with Y. Fyodorov, Y. Gefen, A. Ioselevich, A. Shnirman and M. Skvortsov. The research was funded in part by RFBR Grant No. 14-02-00333, the Council for Grant of the President of Russian Federation (Grant No. MK-4337.2013.2), Dynasty Foundation, RAS Programs ``Quantum mesoscopics and disordered systems'', ``Quantum physics of condensed matter'' and ``Fundamentals of nanotechnology and nanomaterials'', and by Russian Ministry of Education and Science.

\end{acknowledgements}


\appendix

\section{Derivation of $Z(b)$ using the Wei-Norman-Kolokolov transformation \label{WNK-der}}

In this Appendix we present a derivation of the partition function for the Hamiltonian \eqref{ham}. For simplicity, we consider the case of zero magnetic field. We use the notation of Ref. [\onlinecite{BGK2}]. We start from the Hamiltonian $H_0+H_S$. Then the corresponding partition function can be written as
$Z_J=\Tr \langle \exp(-\beta H_S)\rangle$, where $H_S$ is given by Eq. \eqref{Hs} and $\langle \cdots \rangle$ denotes the averaging over all many-particle states with the weight $\exp(-\beta H_0)$. To get rid of terms of the fourth order in electron operators in the exponent $H_S$ we apply the Hubbard-Stratonovich transformation,
\begin{gather}
e^{it[J_{\perp}(S_x^2+S_y^2)+J_z S_z^2]}
 = \lim_{N\rightarrow \infty}\int \Bigl[\prod_{n=1}^{N}d\bm\theta_n\Bigr] \prod_\alpha \mathcal{T}e^{it{\bm\theta_n \bm s_\alpha}/N}\notag \\
 \times  \exp \left [ -\frac{i \Delta}{4}\sum\limits_{n=1}^{N}\left ( \frac{\theta^2_{x,n}+\theta^2_{y,n}}{J_\perp}+\frac{\theta^2_{z,n}}{J_z}\right )\right ] ,
\label{evol}
\end{gather}
where $\Delta=t/N$. Here and further we omit the normalization factors. We restore the correct normalization factor (depending on $T, J_\perp$, and $J_z$) in the final result. To calculate the time-ordered exponent ($\mathcal{T}$) of non-commuting operators it is useful to apply the Wei-Norman-Kolokolov transformation \cite{WeiNorman,Kolokolov} allowing us to rewrite $\mathcal{T}$-exponent as a product of usual exponents:
\begin{align}
& \prod_\alpha \mathcal{T}e^{it{\bm\theta_n \bm s_\alpha}/N}  =
e^{p s_\alpha^{-p} \kappa_{p,N}^p} \exp \left (i s_\alpha^z \Delta \sum_{n=1}^N \rho_{p,n} \right ) \notag \\
& \hspace{1cm} \times
\exp \left ( i  s_\alpha^p \Delta \sum_{n=1}^N \kappa_{p,n}^{-p}  \prod_{j=1}^n e^{- i p \Delta \rho_{p,j}}\right ) ,
\label{eq:HBt}
\end{align}
where $p=\pm$ and $s_\alpha^p = s_\alpha^x+ i p s_\alpha^y$. We employ the initial condition $\kappa_{p,1}^p=0$. The new variables $\rho_p, \kappa_p^p$ and $\kappa_p^{-p}$ are related to the variables $\bm{\theta}$ as follows:
\begin{align}
\frac{\theta_{x,n}-i p \theta_{y,n}}{2}
 & =\kappa^{-p}_{p,n} ,
\,\, \theta_{z,n}  =\rho_{p,n}-\kappa_{p,n}^{-p}(\kappa_{p,n}^p+\kappa_{p,n-1}^p) ,
\notag \\
\frac{\theta_{x,n}+i p \theta_{y,n}}{2}
& =\frac{\kappa^p_{p,n}-\kappa^p_{p,n-1}}{ip\Delta}+\frac{\rho_{p,n}(\kappa_{p,n}^p+\kappa_{p,n-1}^p)}{2}
\notag \\
&  -\frac{(\kappa_{p,n}^p+\kappa_{p,n-1}^p)^2}{4} \kappa_{p,n}^{-p}
 .
\label{WNK}
\end{align}
The vector variables $\bm{\theta}_n$ are real but the transformation \eqref{WNK} assumes that the contour of integration in Eq. \eqref{eq:HBt} has been rotated. In order to preserve the number of variables we impose the following constraints on the new variables: $\rho_{p,n}=-\rho_{p,n}^*$ and $\kappa_{p,n}^+ = (\kappa_{p,n}^-)^*$. We mention that the transformation \eqref{WNK} assumes such a discretization of time that the quantity $(\kappa_{p,N}^p+\kappa_{p,N-1}^p)/2$ corresponds to $\kappa_p^p(t)$ in the continuous limit. In general, there are a lot of discrete representations of $\kappa_p^p(t)$, e.g. of the form $\nu\kappa_{p,N}^p+(1-\nu)\kappa_{p,N-1}^p$ with $0\leqslant \nu \leqslant 1$. However, the choice of the symmetric one (with $\nu=1/2$) is optimal since it allows us to work within the first order in $\Delta$ in Eq. \eqref{evol}. We note that the Jacobian of the transformation \eqref{WNK} is equal to
$\exp(i p \Delta\sum_{n=1}^{N} \rho_{p, n}/2)$.

Having in mind the further usage of the results, we rewrite $\exp(-\beta H)$ as the product $\exp(-it_+ H) \exp(it_- H)$ with $t_+-t_-=-i\beta$. Now, rewriting two exponents in terms of two sets of new variables, we obtain
\begin{widetext}
\begin{align}
Z_J & = \prod_{p=\pm}\Biggl \{ \lim_{N_{p}\rightarrow \infty} \prod_{n_p=1}^{N_p} \int d\kappa_{p,n_p}^p d\kappa_{p,n_p}^{-p} d\rho_{p,n_p} e^{\frac{i p \Delta}{2} \rho_{p,n_p}-\frac{i p \Delta}{4J_z} \rho_{p,n_p}^2-\frac{1}{J_\perp}\kappa_{p,n_p}^{-p}(\kappa_{p,n_p}^p-\kappa_{p,n_p-1}^p)- \frac{i p \Delta\varkappa }{2J_\perp} \rho_{p,n_p}\kappa_{p,n_p}^{-p}(\kappa_{p,n_p}^p+\kappa_{p,n_p-1}^p)}
\notag \\
& \times
e^{ \frac{i p \Delta \varkappa}{4J_\perp} \bigl [\kappa_{p,n_p}^{-p}(\kappa_{p,n_p}^p+\kappa_{p,n_p-1}^p)\bigr ]^2}\Biggr \}  \prod\limits_\alpha
\Tr \prod_{p=\pm} \left [ e^{-ipt_p \epsilon_\alpha n_\alpha} e^{p s^{-p}_\alpha \kappa_{p,N_p}^p}
e^{i s_\alpha^z \Delta \sum_{n=1}^{N_p} \rho_{p,n}}
e^{i  s_\alpha^p \Delta \sum_{n=1}^{N_p} \kappa_{p,n}^{-p}  \exp(- i p \Delta \sum_{j=1}^n \rho_{p,j})} \right ] ,
\label{eq2}
\end{align}
\end{widetext}
where $\varkappa=1-{J_\perp}/{J_z}$ and $\Delta= t_p/N_p$. Let us introduce a set of auxiliary variables $\eta_{p,n_p}$ to get rid of terms of the fourth order in $\kappa_p$'s:
\begin{gather}
e^{\frac{ip \Delta\varkappa}{4J_\perp} [\kappa^{-p}_{p,n_p}(\kappa_{p,n_p}^p+\kappa_{p,n_p-1}^p)]^2}=\int  d\eta_{p,n_p}  e^{ \frac{ip \Delta\varkappa}{4 J_\perp} \eta_{p,n_p}^2} \notag \\
\times  e^{ -\frac{ip \Delta\varkappa}{2 J_\perp} \eta_{p,n_p}\kappa_{p,n_p}^{-p}(\kappa_{p,n_p}^p+\kappa_{p,n_p-1}^p)]} .
\end{gather}
To proceed with the evaluation of $Z_J$ we need to calculate the following integrals over $\kappa_{p}$'s:
\begin{gather}
\prod\limits_{n_p=1}^{N_p}\int d\kappa^p_{p,n_p} d\kappa^{-p}_{p,n_p} \exp\left ( -\frac{1}{J_\perp}\kappa_{p,n_p}^{-p}(\kappa_{p,n_p}^p-\kappa_{p,n_p-1}^p)\right ) \notag \\
\times
\exp \left ( -\frac{i p \Delta\varkappa }{2J_\perp} (\rho_{p,n_p}-\eta_{p,n_p})\kappa_{p,n_p}^{-p}(\kappa_{p,n_p}^p+\kappa_{p,n_p-1}^p) \right ) .
\label{eq:ww1}
\end{gather}
Following Ref. [\onlinecite{Kolokolov}], we introduce the new variables
\begin{equation}
\kappa_{p,n_p}^{-p}=\chi_{p,n_p}^{-p} e^{\alpha_{p,n_p}},\qquad
\kappa_{p,n_p}^{p}=\chi_{p,n_p}^{p} e^{\beta_{p,n_p}} ,
\label{eq:tr2}
\end{equation}
where
\begin{align}
\beta_{p,n_p} & = - i p \Delta \varkappa \sum_{n=1}^{n_p} (\rho_{p,n}-\eta_{p,n}) ,
\notag \\
\alpha_{p,n_p}  & =-\beta_{p,n_p} -\frac{i p\Delta \varkappa}{2}  (\rho_{p,n_p} -\eta_{p,n_p})  .
\label{Jacob}
\end{align}
Such choice of $\alpha_{p,n_p}$ and $\beta_{p,n_p}$ allows us to get rid of terms of the third order (second order in $\chi$'s and first order in $\rho$) in Eq. \eqref{eq:ww1} within the first order in $\Delta$.
The last term in the right hand side of the second equation in \eqref{Jacob} determines the Jacobian $\mathcal{J}_p$ of the  transformation \eqref{eq:tr2}, $\mathcal{J}_p=\exp[ - i p \Delta \varkappa (\rho_{p,n_p} -\eta_{p,n_p})/2]$. We emphasize that it can be missed in the continuous representation.

Evaluating the single-particle trace $\Tr$ in the expression \eqref{eq2} explicitly, one can obtain (the limit $N_{p}\rightarrow \infty$ is assumed)
\begin{widetext}
\begin{gather}
Z_J =  \prod_{p=\pm}\Biggl \{ \prod_{n_p=1}^{N_p} \int d\chi_{p,n_p}^{p} d\chi_{p,n_p}^{-p} d\rho_{p,n_p}d\eta_{p,n_p} \, e^{\frac{i p \Delta}{2} [(1-\varkappa)\rho_{p,n_p}+\varkappa \eta_{p,n_p}]} e^{-\frac{i p \Delta}{4J_z} [\rho_{p,n_p}^2+\frac{\varkappa}{1-\varkappa} \eta_{p,n_p}^2] }
e^{-\frac{1}{J_\perp}\chi_{p,n_p}^{-p}(\chi_{p,n_p}^p-\chi_{p,n_p-1}^p)}
\Biggr \}  \notag \\
\times \prod\limits_\alpha
\Biggl \{1+e^{-2i\epsilon_\alpha(t_+-t_-)}
+2e^{-i\epsilon_\alpha(t_+-t_-)}\cos\left (\frac{\Delta}{2}\sum_{p=\pm}\sum_{n_p=1}^{N_p}\rho_{p,n_p}\right )+ \prod_{p=\pm} e^{-ip \epsilon_\alpha t_p} \exp\left (\frac{i p \Delta}{2} \sum_{n_p=1}^{N_p}\rho_{p,n_p}\right )
\notag \\
\times
\Biggl ( p \chi_{p,N_p}^p e^{- i p \Delta \varkappa \sum_{n_p=1}^{N_p} (\rho_{p,n_p}-\eta_{p,n_p})}
+ i \Delta  \sum_{n_{-p}=1}^{N_{-p}} \chi_{-p,n_{-p}}^p e^{- i p \Delta \varkappa \sum_{n=1}^{n_{-p}} (\rho_{-p,n}-\eta_{-p,n})} e^{i p \Delta \sum_{n=1}^{n_{-p}} \rho_{-p,n}} \Biggr )\Biggr \} .
\end{gather}
 Now the integration over variables $\chi_{p,n_p}$ can be performed (see details in Appendix B of Ref. [\onlinecite{BGK2}]). Then we find
\begin{gather}
Z_J =  \prod_{p=\pm} \Biggl \{\prod_{n_p=1}^{N_p} \int d\rho_{p,n_p} d\eta_{p,n_p}
 \, e^{\frac{i p \Delta}{2} [(1-\varkappa)\rho_{p,n_p}+\varkappa \eta_{p,n_p}]} e^{-\frac{i p \Delta}{4J_z} [\rho_{p,n_p}^2+\frac{\varkappa}{1-\varkappa} \eta_{p,n_p}^2] }\Biggr \}
 \prod_\alpha\left (\oint\limits_{|z_\alpha|=1}\frac{i dz_\alpha}{2\pi z_\alpha^2}\right ) e^{-w}
\notag \\
\times
\exp \Biggl (- 2 v \cos\left [\frac{\Delta}{2}\sum_{p=\pm} \sum_{n_p=1}^{N_p} \rho_{p,n_p}\right ] \Biggr )
\int\limits_0^\infty dy \, e^{-y}
 \exp \Biggl \{ -i J_\perp v y \left ( \prod \limits_{p=\pm} e^{i \frac{p \Delta}{2} \sum_{n_p=1}^{N_p} \rho_{p,n_p}}\right ) \notag \\
\times
 \left ( \sum_{p=\pm} p\, e^{- i p \Delta \varkappa \sum_{n_p=1}^{N_p} (\rho_{p,n_p}-\eta_{p,n_p})}
 \Delta \sum_{n_p=1}^{N_p} e^{- i p \Delta \sum_{n=1}^{n_p} [(1-\varkappa)\rho_{p,n}+\varkappa\eta_{p,n}]}
 \right )
 \Biggr \} .
 \label{eq:Zj1}
\end{gather}
\end{widetext}
Here we introduce the following notation
\begin{gather}
w=\sum_\alpha z_\alpha \left ( 1+e^{-2i\epsilon_\alpha(t_+-t_-)}\right ), \notag \\
v= \sum_\alpha z_\alpha  e^{-i\epsilon_\alpha(t_+-t_-)} .
\label{app:a:bd}
\end{gather}
Let us introduce new variables to make the expression \eqref{eq:Zj1} more standard:
\begin{equation}
\xi_{p}(t)=ip \int_0^t dt^\prime [(1-\varkappa) \rho_{p}(t^\prime)+\varkappa \eta_{p}(t^\prime)]+ \xi_{p}(0) .
\end{equation}
Here we switch to continuous representation. We obtain
\begin{gather}
Z_J =  \prod_\alpha\left (\oint\limits_{|z_\alpha|=1}\frac{i dz_\alpha}{2\pi z_\alpha^2}\right )
\int\limits_0^\infty dy \, e^{-y-w} \prod_{p=\pm} \Biggl \{\int \mathcal{D}[\xi_p,\eta_p] \notag \\
\times
e^{\frac{1}{2}[\xi_p(t_p)-\xi_p(0)]} e^{-\frac{i p}{4J_z} \int_0^{t_p} dt [\frac{(ip \dot{\xi}_p+\varkappa \eta_p)^2}{(1-\varkappa)^2} +\frac{\varkappa \eta_p^2}{1-\varkappa}]}
\Biggr  \}
\notag \\
\times
e^{- 2 v \cosh [\frac{1}{2(1-\varkappa)}\sum_{p=\pm}p  (\xi_{p}(t_p)-\xi_p(0) - i p \varkappa\int_0^{t_p} dt^\prime \eta_p(t^\prime) ) ]} \notag \\
\times
 \exp \Biggl \{ -i J_\perp v y \left ( \prod \limits_{p=\pm} e^{\frac{1}{2(1-\varkappa)} [\xi_p(t_p)-\xi_p(0)-ip\varkappa \int_0^{t_p} dt \eta_p(t)]
} \right ) \notag \\
\times
 \left ( \sum_{p=\pm} p\, e^{\frac{1}{1-\varkappa}
 [\xi_p(0)-\varkappa \xi_p(t_p)+ip \varkappa \int_0^{t_p} dt \eta_p(t)]}
\int\limits_0^{t_p}\!\! dt \, e^{- \xi_p(t)}
 \right )
 \Biggr \} .
 \label{eq:Zj2}
\end{gather}
There is some freedom in choosing the initial conditions for field variables $\xi_p(t)$. It is convenient to choose them such that the following relations hold:
\begin{gather}
\sum_{p=\pm} \xi_p(t_p) + 2 \ln (4v y) = 0 ,\\
\sum_{p=\pm} p \Bigl [ \xi_p(0)-\varkappa \xi_p(t_p) + i p \varkappa \int\limits_0^{t_p}dt \eta_p(t) \Bigr ] = 0 .
\label{Eq:BC}
\end{gather}
Then Eq. \eqref{eq:Zj2} can be rewritten as
\begin{gather}
Z_J = \prod_\alpha\left (\oint\limits_{|z_\alpha|=1}\frac{i dz_\alpha}{2\pi z_\alpha^2}\right ) \int\limits_0^\infty dy \, e^{-y-w}
 \prod_{p=\pm}\Biggl \{ \int \mathcal{D}[\xi_p,\eta_p] \notag \\
\times
\int\limits_{-\infty}^\infty dx\,  e^{\frac{1}{2}[\xi_p(t_p)-\xi_p(0)]}
e^{-\frac{i p}{4J_z} \int_0^{t_p} dt [\frac{(ip \dot{\xi}_p+\varkappa \eta_p)^2}{(1-\varkappa)^2} +\frac{\varkappa \eta_p^2}{1-\varkappa}]} \notag \\
\times
e^{\frac{i x p}{1-\varkappa} [ \xi_p(0)-\varkappa \xi_p(t_p) + i p \varkappa \int_0^{t_p}dt \eta_p(t)]}
\Biggr \} e^{- 2 v \cosh\frac{\xi_{+}(t_+)-\xi_{-}(t_-)}{2}}
\notag \\
\times
 e^{-\frac{i J_\perp}{4} \sum_{p=\pm} p \int_0^{t_p} dt\,  e^{- \xi_p(t)}}
\delta \left ( \sum_{p=\pm} \xi_p(t_p) + 2 \ln (4v y)\right ) .
 \label{eq:Zj3}
\end{gather}
Integrating over the variables $\eta_p$ we find
\begin{gather}
Z_J = \prod_\alpha\left (\oint\limits_{|z_\alpha|=1}\frac{i dz_\alpha}{2\pi z_\alpha^2}\right )
\int\limits_{-\infty}^\infty dx  \prod_{p=\pm}  \Biggl \{ e^{-i J_z \varkappa  x^2  p t_p}  \notag \\
\times \int \mathcal{D}[\xi_p] \,e^{i p \int_0^{t_p} dt \mathcal{L}_p} e^{-(1-2i p x) \xi_p(0)/2}
\Biggr \}
 \notag \\
 \times
\int\limits_0^\infty \frac{dy}{4 y v} \, e^{-y-w} \,
\exp \left ( - 2 v \cosh\Bigl [\frac{1}{2}\sum_{p=\pm}p \xi_{p}(t_p) \Bigr] \right )
\notag \\
\times \delta \left ( \sum_{p=\pm} \xi_p(t_p) + 2 \ln (4v y)\right ) .
 \label{eq:Zj4}
\end{gather}
The functional integral \eqref{eq:Zj4} is of Feynman-Kac type with the Lagrangian
\begin{equation}
\mathcal{L}_p = \frac{1}{4J_\perp} \dot\xi_p^2 - \frac{J_\perp}{4} e^{-\xi_p} .
\end{equation}
Then the calculation of the partition function can be reduced to an evaluation of two matrix elements:
\begin{gather}
Z_J = \prod_\alpha\left (\oint\limits_{|z_\alpha|=1}\frac{i dz_\alpha}{2\pi z_\alpha^2}\right )
\int\limits_{-\infty}^\infty dx   e^{-i J_z \varkappa  x^2  (t_+-t_-)} \notag \\
\times
\int\limits_0^\infty \frac{dy}{4 y v} \, e^{-y-w} \prod_{p=\pm}  \left \{ \int d\xi_p d\xi_p^\prime \,
\, e^{-(1-2i p  x) \xi_p^\prime/2}
\right \}
\notag \\
\times
\delta \left ( \sum_{p=\pm} \xi_p + 2 \ln (4v y)\right )
e^{- 2 v \cosh [(\xi_+-\xi_-)/2]}
\notag \\
\times
\langle \xi_+ | e^{-i \mathcal{H}_J t_+} | \xi_+^\prime \rangle
\langle \xi_-^\prime | e^{i \mathcal{H}_J t_-} | \xi_- \rangle  .
 \label{eq:Zj5}
\end{gather}
Here the one-dimensional quantum mechanical Hamiltonian
\begin{equation}
\mathcal{H}_J=-J_\perp\frac{\partial^2}{\partial\xi^2}+\frac{J_\perp}{4}e^{-\xi} .
\end{equation}
Its eigenfunctions are given by the modified Bessel functions $K_{2i\nu}$ where $\nu$ is a real number:
\begin{equation}
\langle\xi |\nu\rangle = \frac{2}{\pi} \sqrt{\nu \sinh(2\pi \nu)} K_{2i\nu}(e^{-\xi/2}) .
\end{equation}
The eigenvalues of $\mathcal{H}_J$ are equal to $J \nu^2$: $\mathcal{H}_J |\nu\rangle = J\nu^2 |\nu\rangle$.
After integration over $y$ we obtain
\begin{align}
Z_J & = \frac{4}{\pi^2}\prod_\alpha\left (\oint\limits_{|z_\alpha|=1}\frac{i dz_\alpha}{2\pi z_\alpha^2}\right )
\int\limits_{-\infty}^\infty dx   e^{-i J_z \varkappa x^2  (t_+-t_-)} \notag \\
& \times
\int\limits_0^\infty \frac{d\nu}{v} \, \nu \sinh(2\pi \nu) K_{2i\nu}(2v) e^{-w}
\notag \\
& \times
\prod_{p=\pm}  \left \{ \int d\xi_p d\xi_p^\prime \,
\, e^{-(1-2i p x) \xi_p^\prime/2} K_{2i\nu}(e^{-\xi_p/2})
\right \}
\notag \\
& \times
\langle \xi_+ | e^{-i \mathcal{H}_J t_+} | \xi_+^\prime \rangle
\langle \xi_-^\prime | e^{i \mathcal{H}_J t_-} | \xi_- \rangle  .
 \label{eq:Zj6}
\end{align}
Here we use the following result (see formula 6.794.11 on p. 794 of Ref. [\onlinecite{GR}])
\begin{gather}
\int\limits_0^\infty d\nu \, \nu \sinh(2\pi \nu) K_{2i\nu}(2v)K_{2i\nu}(e^{-\xi_+/2}) K_{2i\nu}(e^{-\xi_-/2})
\notag \\
= \frac{\pi^2}{16}\exp \left ( -\frac{1}{4v} e^{-\frac{\xi_++\xi_-}{2}} -2 v \cosh\frac{\xi_+-\xi_-}{2}
\right ) .
\end{gather}
Integration over $\xi_p$ can be now easily performed, and we obtain
\begin{gather}
Z_J = \frac{32}{\pi^2}\prod_\alpha\left (\oint\limits_{|z_\alpha|=1}\frac{i dz_\alpha}{2\pi z_\alpha^2}\right )
\int\limits_{-\infty}^\infty dx   e^{-i J_z \varkappa x^2  (t_+-t_-)} \int\limits_0^\infty \frac{d\nu}{v}  \notag \\
\times
 \nu \sinh(2\pi \nu) K_{2i\nu}(2v) e^{-w}
 e^{-i J_\perp \nu^2 (t_+-t_-)} \int d\eta_+d\eta_-
\notag \\
\times
\, e^{-2\eta_++4i x \eta_-} K_{2i\nu}(e^{-\eta_{+}-\eta_-})
 K_{2i\nu}(e^{-\eta_{+}+\eta_-})
 .
 \label{eq:Zj7}
\end{gather}
Using the identity  (see formula 6.521.3 on p. 658 of Ref. [\onlinecite{GR}])
\begin{equation}
\int\limits^\infty_0 dx\, x K_\nu(ax)K_\nu(bx)=\frac{\pi(ab)^{-\nu}(a^{2\nu}-b^{2\nu})}{2\sin(\pi\nu)(a^2-b^2)},
\end{equation}
we can perform the integration over $\eta_+$. With the help of the integral representation of the modified Bessel function
\begin{equation}
K_\nu(x)=\int\limits_0^\infty dh e^{-x \cosh h + \nu h},
\end{equation}
we integrate over $\nu$. Finally, integration over $x$ yields
\begin{gather}
Z_J = \frac{2 e^{-\beta J_\perp/2}}{\pi \beta \sqrt{J_\perp (J_z-J_\perp)}}
\int\limits_{-\infty}^\infty dh \, \sinh h
\prod_{\alpha,\sigma} \left (1+e^{-\epsilon_\alpha\beta+h\sigma}\right )
 \notag \\
 \times
 e^{-\frac{h^2}{J_\perp\beta}} \int\limits_{-\infty}^\infty d\eta_-
\, e^{-\frac{4\eta_-^2}{\varkappa J_\perp \beta}}
\frac{\sinh \frac{4\eta_- h}{J_\perp \beta}}{\sinh (2\eta_-)}
 .
 \label{eq:Zj8}
\end{gather}
Here we restored the correct numerical factor using the normalization condition $Z_J=1$ at $\epsilon_\alpha\to +\infty$.

In order to derive the partition function for the Hamiltonian \eqref{ham} from Eq. \eqref{eq:Zj8} one needs to make the substitution $\epsilon_\alpha \to \epsilon_\alpha + i \phi_0 T$ and to integrate over the variable $\phi_0$:
\begin{equation}
Z =
\sum_{k\in \mathbb{Z}} e^{-\beta E_c(k-N_0)^2} \
\int\limits_{-\pi}^{\pi}\frac{d\phi_0}{2\pi}  e^{i\phi_0 k}
Z_J
\end{equation}
Then we obtain Eq. \eqref{IR} with $b=0$.

It is easy to obtain the partition function with non-zero magnetic field. The field shifts the $z$ projection of the total spin   in the evolution operator
\eqref{evol}: $S_z\rightarrow S_z+ \frac{B}{2 J_z}$. This shift affects only the boundary conditions on $\xi$ in \eqref{Eq:BC}.

\section{Asymptotic results for the functions $F_1(x,y)$ and $F_2(x,y)$ \label{App:F1:ZZF1}}

At $y\ll \min\{1,1/\sqrt{x}\}$, the value of the integral in Eq. \eqref{eq:F1} is determined by the region $||t|-xy/2|\sim 1$. Thus one can expand $\sinh$ in the denominator into a series in $yt\sim y^2x\ll 1$ and obtain
\begin{equation}
F_1(x,y) = \frac{\sqrt\pi}{y} \erfi(x y/2) - \frac{x y^2}{12} \exp(x^2y^2/4) .
\label{eq:App:F1:ZZF1r1}
\end{equation}
Here we performed the expansion to the second order in $y t$, having the further calculation for the spin susceptibility in mind.

At $1/\sqrt{x}\ll y \ll 1$, the argument of the $\sinh$ in the denominator is large and one can make the following replacement: $\sinh yt \sim \sgn(t) \, \exp(y|t|)/2$. Then we find
\begin{equation}
F_1(x,y) = 2\exp\Bigl [(x-1)^2y^2/4\Bigr ] .
\label{eq:App:F1:ZZF1r2}
\end{equation}
The same simplification for $\sinh yt$ can be used in the limit $y\gg 1$ and $x\geqslant 1$. Then we obtain
\begin{equation}
F_1(x,y) = e^{(x-1)^2y^2/4} \left [ 1 + \erf\bigl ((x-1)y/2\bigr) \right ] .
\label{eq:App:F1:ZZF1r3}
\end{equation}
At $y\gg 1$ and $x\ll 1$ the relevant region of integration in Eq. \eqref{eq:F1} is determined by the denominator. Thus one can omit $e^{-t^2}$, expand $\sinh (xyt)$ in the numerator and find
\begin{equation}
F_1(x,y)=\frac{\pi^{3/2} x}{2 y} .
\label{eq:App:F1:ZZF1r4}
\end{equation}

The only relevant case for our values of $x$ and $y$ in Eq. \eqref{eq:F2} is the case with $y\ll 1/\sqrt{x}$. In this regime  the denominator can be substituted by $yt$ and the region of integration can be extended to infinity. Then the following result can be obtained from \eqref{eq:App:F1:ZZF1r1} by replacement $y\rightarrow i y$:
\begin{equation}
F_2(x,y) = \frac{\sqrt\pi}{y} \erf(x y/2) + \frac{x y^2}{12} \exp(-x^2y^2/4)  .
\label{eqnF2asymp}
\end{equation}

\section{Derivation of Eqs. \eqref{EQm2} and \eqref{EQm2:gen} \label{App:ZnZn}}

In this Appendix we present brief arguments why Eqs. \eqref{EQm2} and \eqref{EQm2:gen} are correct. Since
Eq. \eqref{EQm2} can be obtained from Eq. \eqref{EQm2:gen}, we consider only the latter. We start from the following exact expression
\begin{align}
Z_{n_\uparrow}Z_{n_\downarrow}  = \int \limits_{-\pi}^\pi \frac{d\theta_1 d\theta_2}{(2\pi)^2} e^{-i\phi n
-i\theta m} \prod\limits_{\sigma=\pm} e^{-\beta \Omega_0(i T \phi + i \sigma T \theta/2)}
\end{align}
where $\theta_{1,2} = \phi \pm \theta/2$. As usual, at $\delta\ll T$ the integral over $\phi$ can be performed in the saddle-point approximation. This yields
\begin{align}
Z_{n_\uparrow}Z_{n_\downarrow}  \approx \sqrt{\frac{\beta \delta}{4\pi}}
e^{-\beta\mu_n n - 2 \beta \Omega_0 (\mu_n)} \mathcal{X}_m(\beta\delta) ,
\end{align}
where
\begin{align}
\mathcal{X}_m(x)=\int\limits_{-\pi}^\pi \frac{d\theta}{\pi} e^{-2 m i \theta} e^{-\theta^2/x - V(i\theta)} .
\end{align}
The function $\mathcal{X}_m(x)$ can be rewritten as
\begin{align}
\mathcal{X}_m(x) = \int\limits_{-\infty}^\infty \frac{d\theta}{\pi} e^{-\theta^2/x- V(i\theta)} \cos{2m \theta}
+ \mathcal{R}_m(x) .
\end{align}
Now we bound
\begin{equation}
\mathcal{R}_m(x) =2 \int\limits_{\pi}^\infty \frac{d\theta}{\pi} e^{-\theta^2/x- V(i\theta)} \cos{2m \theta}
\end{equation}
from above. Using that random function $V(i\theta)$ depends, in fact, on $\sin^2(\theta/2)$ (cf. Eq. \eqref{EQ:APP:G}), we obtain the following set of inequalities:
\begin{align}
\bigl | \mathcal{R}_m(x) \bigr | & \leqslant 4 \int\limits_{0}^{2\pi} \frac{d\theta}{2\pi} \sum_{l=0}^\infty
e^{-(\theta+\pi +2\pi l)^2/x} e^{-V(i\theta)} \notag \\
& \leqslant  4  \sum_{l=0}^\infty
e^{-\pi^2(2l+1)^2/x}  \int\limits_{0}^{2\pi} \frac{d\theta}{2\pi}  e^{-V(i\theta)}
\notag \\
& \leqslant  4  \sum_{l=1}^\infty
e^{-\pi^2 l /x}  \int\limits_{0}^{2\pi} \frac{d\theta}{2\pi}  e^{-V(i\theta)}
\notag \\
& \leqslant  \frac{4}{e^{\pi^2/x}-1} \int\limits_{0}^{2\pi} \frac{d\theta}{2\pi}  e^{-V(i\theta)} .
\end{align}
Hence we demonstrate that $\bigl | \mathcal{R}_m(x) \bigr | \leqslant O(e^{-\pi^2/x})$ is independent of $m$.  Therefore we can write $\mathcal{X}_m(x)$ at $x\ll 1$ as follows:
\begin{align}
\mathcal{X}_m(x) \approx \int\limits_{-\infty}^\infty \frac{d\theta}{\pi} e^{-\theta^2/x- V(i\theta)} \cos{2m \theta} .
\end{align}

\section{Correlation function $V(h_1)V(h_2)$ \label{app:Lh}
}

In this Appendix we present a brief derivation of Eq. \eqref{corrVV}. The correlation function of the single-particle density of states is given by~\cite{Mehta}
\begin{equation}
\langle \delta\nu_0(E)\delta\nu_0(E+\omega)\rangle = \frac{1}{\delta^2} \left [\delta\left (\frac{\omega}{\delta}\right )- R\left (\frac{\pi \omega}{\delta}\right )\right ] .
\label{App_dnu1}
\end{equation}
Here the function $R(x)$ depends on the statistics of the ensemble of single-particle energies. Using Eq. \eqref{App_dnu1}, the identity $\int_{-\infty}^\infty R(x) dx=\pi$ and the definition of $V(h)$ we obtain
\begin{align}
\overline{V(h_1)V(h_2)}  & = T^2 \int_{-\infty}^\infty \frac{dE d\omega}{\delta^2} R\left (\frac{\pi T \omega}{\delta}\right ) \Bigl [ g(E,h_1) g(E,h_2) \notag \\
& - g(E+\omega/2,h_1) g(E-\omega/2,h_2)\Bigr ] ,\label{eqC_app}
\end{align}
where
\begin{equation}
g(E,h) = \ln \left [ 1+\frac{\sinh^2(\frac{h}{2})}{\cosh^2(\frac{E}{2})}\right ] .
\label{EQ:APP:G}
\end{equation}
The function $g(E,h)$ has the following Fourier transform with respect to variable $E$:
\begin{align}
g(t,h) & = \int_{-\infty}^\infty \frac{dE}{2\pi}\, e^{i E t} \, g(E,h) = \frac{1}{2\pi t} \Imag \int_{-\infty}^\infty dE e^{i E t} \notag \\
& \times \tanh\frac{E}{2} \frac{\sinh^2(h/2)}{\sinh^2(h/2)+\cosh^2(E/2)} .
\label{eqgth}
\end{align}
Since the function $g(E,h)$ is even in $E$, the function $g(t,h)$ is even in $t$. The function under the integral in the r.h.s. of Eq. \eqref{eqgth} has poles at $E=\pi (2n+1)i, \pm h + \pi(2m+1) i$ where $n$ and $m$ are integers. Computation of the residues yields
\begin{align}
g(t,h) & =  \frac{1}{2\pi t} \Imag 4\pi i \sum_{n\geqslant 0} e^{-\pi(2n+1) t} \left ( 1- \frac{1}{2} e^{-i ht} -\frac{1}{2} e^{i h t}\right )\notag \\
&  \hspace{2cm} = \frac{1-\cos(ht)}{t \sinh(\pi t)} .
\end{align}
Substitution into Eq. \eqref{eqC_app} leads to
\begin{align}
\overline{V(h_1)V(h_2)}  & = 2\pi T^2 \int_{-\infty}^\infty \frac{dt d\omega}{\delta^2} R\left (\frac{\pi T \omega}{\delta}\right ) g(t,h_1) \notag \\
& \times g(t,h_2) \Bigl [ 1- e^{-i\omega t} \Bigr ] .
\end{align}
At $x\gg 1$ the function $R(x)$ has the following asymptotic behavior:
\begin{equation}
R(x) = \frac{1}{\bm{\beta} x^2} , \qquad x\gg 1 .
\end{equation}
Recall that $\bm{\beta}=1$ for the orthogonal Wigner-Dyson ensemble, $\bm{\beta}=2$ for the unitary Wigner-Dyson ensemble and $\bm{\beta}=4$ for the symplectic Wigner-Dyson ensemble. Then at $\max\{|h|,T/\delta\}\gg 1$ we find
\begin{align}
\overline{V(h_1)V(h_2)}  & = \frac{4}{\bm{\beta}} \int_{0}^\infty dt\, \frac{[1-\cos(h_1 t)][1-\cos(h_2 t)]}{t \sinh^2(\pi t)} \notag \\ & =  \sum_{\sigma=\pm} L(h_1+\sigma h_2)- 2 L(h_1) - 2L(h_2) ,
\end{align}
where
\begin{equation}
L(h) = \frac{2}{\bm{\beta}} \int_{0}^\infty dt\, \frac{\cos(h t)-1+h^2t^2/2}{t \sinh^2(\pi t)}
\end{equation}
is even in $h$. Next, for $h>0$
\begin{align}
L^{\prime}(h) & = \frac{2}{\bm{\beta}} \int_{0}^\infty dt\, \frac{ht -\sin(h t)}{\sinh^2(\pi t)} \notag \\
& = \frac{8}{\bm{\beta}} \int_{0}^\infty dt\, \sum_{n=1}^\infty n [ht -\sin(h t)] e^{-2\pi n t} \notag \\
& = \frac{2 h}{\pi^2\bm{\beta}} \left [ \Real \psi\left (1+\frac{i h}{2\pi}\right ) - \psi(1)\right ].
\end{align}
This is the Eq. \eqref{corrVV} of the paper. Using the well-known  asymptotic expressions for the Euler digamma function $\psi(x)$ at small and large values of its argument one arrives at Eq. \eqref{Sassymp}.

\section{Forth order perturbation theory for $\overline{\chi}_{zz}$ in the case of the Ising exchange\label{app:2dOrder}}

In this Appendix we present the derivation of the perturbative results \eqref{eq:chizz:regI} and \eqref{eq:chizz:regII} for $b=0$. In addition, we compute the next order in $L$ for the correction to $\overline{\chi}_{zz}$.

We start from the expansion of the average $\ln Z_S$ to the fourth order in $V$:
\begin{align}
\overline{\ln Z_S} & = \frac{1}{2} \ln \frac{\bar{J}_z}{J_z} - \frac{1}{2} F_{2} - \frac{1}{2} F_{1,1} - \frac{1}{24} F_4 -\frac{1}{8} F_{2,2} - \frac{1}{6} F_{3,1} \notag \\
& - \frac{1}{2} F_{2,1,1} - \frac{1}{4} F_{1,1,1,1} + O(V^6) .
\end{align}
Here we introduced
\begin{align}
F_{k_1,\dots,k_q} & = (-1)^q \int_{-\infty}^\infty \frac{dh_1\dots dh_q}{\pi^{q/2}} \, \exp \left (\sum_{j=1}^q h_j^2\right ) \notag \\
& \times \overline{V^{k_1}(h_1)\dots V^{k_q}(h_q)} .
\end{align}

\subsection{Second order in $V$}

The contribution of the second order in $V$ is given by $F_2$ and $F_{1,1}$. We find
\begin{equation}
F_2+F_{1,1} = 2\int_0^\infty \frac{d h}{\sqrt{\pi}} \, e^{-h^2} \left [ 2 L\left (h \sqrt{2y}\right )- L\left (2h \sqrt{y}\right ) \right ] .
\label{eq:F2_11}
\end{equation}
Here we remind $y=\beta \bar{J}_z$. It is instructive to compare the second order contribution \eqref{eq:F2_11} with the second order contribution to the variance of $\ln Z_S$:
\begin{align}
\overline{\bigl (\ln Z_S - \overline{\ln Z_S}\bigr )^2} = F_{1,1} & = 4\int_0^\infty \frac{d h}{\sqrt{\pi}} \, e^{-h^2} \Bigl [ L\left (h \sqrt{2 y}\right ) \notag \\
& - 2 L\left (h \sqrt{y}\right ) \Bigr ] .
\label{eq:Var}
\end{align}

In the regime $T\gg \bar{J}_z$ the arguments of $L$ in the right hand side of Eqs. \eqref{eq:F2_11} and \eqref{eq:Var} are small. Using the asymptotic expression for $L(h)$ at $|h|\ll 1$, we obtain
\begin{equation}
F_2+F_{1,1} =- \frac{3\zeta(3)}{4\pi^4 \bm{\beta}}\frac{\bar{J}_z^2}{T^2} , \qquad F_{1,1} = \frac{3\zeta(3)}{8\pi^4 \bm{\beta}}\frac{\bar{J}_z^2}{T^2} .
\label{eq:F2+11small}
\end{equation}
The result \eqref{eq:F2+11small} for $F_2+F_{1,1}$ is translated into Eq. \eqref{eq:chizz:regI} of the paper. From Eq. \eqref{eq:F2+11small} we find that
\begin{equation}
\frac{\overline{\bigl (\chi_{zz}-\overline{\chi}_{zz}\bigr)^2}}{\overline{\chi}_{zz}^2} \propto \frac{\bar{J}_z^2}{\pi^2\bm{\beta}T^2} \ll 1, \qquad T\gg \bar{J}_z .
\end{equation}

At low temperatures $T\ll \bar{J}_z$ the asymptotic expression of $L(h)$ for $|h|\gg 1$ must be used in Eq. \eqref{eq:F2_11}. We find
\begin{equation}
F_2+F_{1,1} =- \frac{\ln 2}{\pi^2\bm{\beta}}\frac{\bar{J}_z}{T} , \qquad F_{1,1} = \frac{\ln 2}{\pi^2\bm{\beta}}\frac{\bar{J}_z}{T} .
\label{eq:F2F11}
\end{equation}
From Eq. \eqref{eq:F2+11small} it follows that
\begin{equation}
\frac{\overline{\bigl (\chi_{zz}-\overline{\chi}_{zz}\bigr)^2}}{\overline{\chi}_{zz}^2} \propto \frac{\bar{J}_z}{\pi^2\bm{\beta}T} \ll 1, \qquad \frac{\bar{J}_z }{\pi^2\bm{\beta}}\ll T\ll \bar{J}_z .
\label{eq:Var2}
\end{equation}
In view of the result \eqref{eq:Var2} we can expect that $\ln Z_S$ has a normal distribution with mean
$[\ln (\bar{J_z}/J_z) -F_2-F_{1,1}]/2$ and variance $F_{1,1}$ in the regime ${\bar{J}_z }/{(\pi^2\bm{\beta})}\ll T\ll \bar{J}_z$. For $T=3\delta$ and $J_z/\delta= 0.97$ the complementary cumulative distribution function for the normal distribution and the complementary cumulative distribution function obtained numerically for the process $V(h)$ are compared in Fig. \ref{Fig:CCDF}. We note that for $T=3\delta$ and $J_z/\delta= 0.94$ numerical integration of  Eqs. \eqref{eq:F2_11} and \eqref{eq:Var} yields $F_2+F_{1,1}\approx -0.07$ and $F_{1,1}\approx 0.05$. These values are still different from the asymptotic estimates \eqref{eq:F2F11}.

\subsection{Fourth order in $V$}

In the regime $T\gg \bar{J}_z$ the fourth order contributions are proportional to
$(J_z/T)^4$ and therefore negligible. For low temperatures $T\ll \bar{J}_z$ the contributions of the fourth order in $V$ are listed below:
\begin{equation}
F_4 = -3 \int_{-\infty}^\infty \frac{dh}{\sqrt{\pi}} \, e^{-h^2} \Bigl [ \overline{V^2(h\sqrt{y})}\Bigr ]^2 = -36 \ln^2 2 \, z^4 ,
\end{equation}
\begin{align}
F_{2,2} & = \left [ \int_{-\infty}^\infty \frac{dh}{\sqrt{\pi}} \, e^{-h^2}  \overline{V^2(h\sqrt{y})} \right ]^2 \notag \\ & +
2  \int_{-\infty}^\infty \frac{dh_1d h_2}{\pi} \, e^{-h_1^2-h^2_2} \Bigl [ \overline{V(h_1\sqrt{y})V(h_2\sqrt{y})}\Bigr ]^2
\notag \\
& =  \left ( 4 \ln^2 2 +  8 b_{2,2} \right )\,  z^4 , \\
b_{2,2} & = \frac{1}{2} \int_0^{2\pi} \frac{d\phi}{2\pi} \Bigl ( \overline{v(\cos\phi)v(\sin\phi)}\Bigr )^2 \approx 0.35 ,
\end{align}
\begin{align}
F_{3,1} & = 3 \int_{-\infty}^\infty \frac{dh_1d h_2}{\pi} \, e^{-h_1^2-h^2_2} \overline{V(h_1\sqrt{y})V(h_2\sqrt{y})}\notag \\
& \times \overline{V^2(h_2\sqrt{y})} = 12 \ln^2 2 \, z^4 ,
\end{align}
\begin{align}
F_{2,1,1} = & - \int_{-\infty}^\infty \frac{dh_1d h_2dh_3}{\pi^{3/2}} \, e^{-h_1^2-h^2_2-h_3^2}\,\,
  \overline{V(h_1\sqrt{y})V(h_2\sqrt{y})} \notag \\
  & \times \Bigl [ \overline{V^2(h_3\sqrt{y})} + 2 \overline{V(h_1\sqrt{y})V(h_3\sqrt{y})}
\Bigr ] \notag \\
= & - \left (2\ln^2 2 +  2 b_{2,1,1} \right )\,  z^4 ,
\\
b_{2,1,1} = & \frac{15}{4} \int_0^{2\pi}\frac{d\phi}{4\pi}\int_0^\pi d\theta \, \sin^3\theta \,\,
 \overline{v(\cos\phi)v(\sin\phi)}\notag \\
 & \times \overline{v(\cos\theta)v(\sin\theta\cos\phi)}\approx 0.79 ,
\end{align}
\begin{equation}
F_{1,1,1,1} = 3 \left [ \int_{-\infty}^\infty \frac{dh}{\sqrt{\pi}} \, e^{-h^2}  \overline{V^2(h\sqrt{y})} \right ]^2
= 3 \ln^2 2 \, z^4 .
\end{equation}
Here we recall that $z^2 =  \bar{J}_z/(\pi^2 \bm{\beta} T)$.
Summing up, for $T\ll \bar{J}_z$ we obtain
\begin{equation}
\overline{\ln Z_S} = \frac{1}{2} \ln \frac{\bar{J_z}}{J_z} + \frac{\ln 2}{2\pi^2\bm{\beta}}\frac{\bar{J}_z}{T} + \frac{a_2}{4} \left (\frac{\bar{J}_z}{\pi^2\bm{\beta} T}\right )^2 ,
\label{eq:lnZs2d}
\end{equation}
where
\begin{equation}
a_2 = -3\ln^2 2 - 4 b_{2,2} + 4 b_{2,1,1} \approx 0.29 .
\end{equation}
Using Eq. \eqref{eq:lnZs2d} and the definition of the spin susceptibility we obtain for $b=0$
\begin{equation}
\overline{\chi}_{zz} = \frac{1}{2(\delta -J_z)} \left [ 1 + \frac{\bar{J}_z \ln 2}{\pi^2\bm{\beta} T}{T} + a_2 \left (\frac{\bar{J}_z}{\pi^2\bm{\beta} T}\right )^2 \right ] .
\end{equation}

\end{document}